\documentclass[aps,prd,onecolumn,superscriptaddress,nofootinbib,floatfix,onecolumn]{revtex4}

\usepackage{dcolumn}
\usepackage{bm}

\usepackage[dvips]{graphicx}
\usepackage{float}
\usepackage{epsfig}
\usepackage{graphicx}
\usepackage{longtable}
\usepackage{rotating}
\usepackage{ulem}
\usepackage{subfigure}
\usepackage{amsmath,amssymb,color,hyperref}
\usepackage{mathrsfs}
\usepackage{calrsfs}
\usepackage{txfonts}
\usepackage{pslatex}
\usepackage{color}
\usepackage{mathtools}
\usepackage{tabularx}
\usepackage{booktabs}
\usepackage{xspace}

\def\ttt#1{\texttt{\small #1}}

\newcommand{\conex}{\textsc{conex}}
\newcommand{\corsika}{\textsc{corsika}}
\newcommand{\seneca}{\textsc{seneca}}
\newcommand{\pythia}{\textsc{pythia}}
\newcommand{\epos}{\textsc{epos}}
\newcommand{\eposlhc}{\textsc{epos-lhc}}
\newcommand{\qgsjet}{\textsc{qgsjet}} 
\newcommand{\qgsjetII}{\textsc{qgsjet-ii-04}} 
\newcommand{\sibyll}{\textsc{sibyll}}

\newcommand{\sqrts}{\sqrt{s}}

\newcommand{\pp}{p-p}
\newcommand{\epem}{e^+e^-}
\newcommand{\alphas}{\alpha_{\rm s}}

\newcommand{\pT}{\rm p_{_{\rm T}}}
\newcommand{\ECR}{\rm E_{_{\rm CR}}}
\newcommand{\Xmax}{\left<\rm X_{\rm max}\right>}

\newcommand{\smax}{\sigma_{\rm X_{max}}}

\newcommand{\Pom}{\mathbb{P}}

\newcommand{\meanpt}{\rm \left< p_{\rm T} \right>}

\newcommand*{\eg}{e.g.,\@\xspace}
\newcommand*{\ie}{i.e.,\@\xspace}
\newcommand*{\cm}{c.m.\@\xspace}
\newcommand*{\etc}{etc.\@\xspace}
\newcommand*{\elm}{e.m.\@\xspace}

\begin{document}

\title{Impact of QCD jets and heavy-quark production in cosmic-ray proton\\ atmospheric showers up to 10$^{20}$~eV}
 
\author{David~d'Enterria}
\affiliation{CERN, EP Department, 1211 Geneva, Switzerland}
\author{Tanguy Pierog}
\affiliation{Karlsruhe Institute of Technology (KIT), IKP, Postfach 3640, 76021 Karlsruhe, Germany}
\author{Guanhao Sun}
\affiliation{CERN, EP Department, 1211 Geneva, Switzerland}
\affiliation{Hong Kong Univ.\,of Science and Technology, HKUST, Hong Kong, China}
\affiliation{Center for Theoretical Physics and Dept. of Physics, Columbia Univ., New York, NY 10027, USA}

\begin{abstract}
\noindent 
The \pythia~6 Monte Carlo (MC) event generator, commonly used in collider physics, is interfaced 
for the first time with a fast transport simulation of a 
hydrogen atmosphere, with the same density as air, in order to study the properties of extended atmospheric 
showers (EAS) produced by cosmic ray protons with energies $\ECR\approx 10^{14}$--$10^{20}$~eV.
At variance with the hadronic MC generators (\eposlhc, \qgsjet, and \sibyll) commonly used in cosmic-rays physics, 
\pythia\ includes the generation of harder hadronic jets and heavy 
(charm and bottom) quarks, thereby producing higher transverse momentum final particles, that could explain 
several anomalies observed in the data. 
The electromagnetic, hadronic, and muonic properties of EAS generated with various settings of \pythia~6, 
tuned to proton-proton data measured at the LHC, are compared to those from \eposlhc, \qgsjet~01, \qgsjetII, and \sibyll~2.1.
Despite their different underlying parton dynamics, the characteristics of the EAS generated with \pythia~6 
are in between those predicted by the rest of MC generators. The only exceptions are the muonic 
components at large transverse distances from the shower axis, where \pythia\ predicts more activity
than the rest of the models. Heavy-quark production, as implemented in this study for a hydrogen atmosphere, does
not seem to play a key role in the EAS muon properties, pointing to nuclear effects as responsible of
the muon anomalies observed in the air-shower data.
\end{abstract}


\maketitle

\section{Introduction}

Ultrahigh-energy cosmic rays (UHECR) with energies up to $\ECR\approx 10^{20}$~eV 
are the most energetic particles known in the universe. Their exact nature and origin, protons 
or heavier ions accelerated in various extreme extragalactic environments, remain still open questions 
today (see \eg~\cite{Mollerach:2017idb,Dawson:2017rsp} for recent reviews). The flux of UHECR impinging
on earth is very scarce (less than 1 particle per km$^2$ per century at the highest energies),
and their detection is only possible through the huge extensive air showers (EAS) of secondary particles 
that they produce in electromagnetic and hadronic interactions with the nitrogen and oxygen nuclei in
the atmosphere. These showers include an electromagnetic (\elm) component consisting of electrons, positrons, 
and photons (mostly coming from the $\pi^0\to\gamma\gamma$ decays of the produced neutral pions);
a hadronic component mostly consisting of protons, neutrons, and charged pions and kaons; 
as well as muonic and neutrino components (mostly issuing from the 
hadronic shower, via decays of charged pions  and kaons). As the cascade develops in the atmosphere, the number of 
particles in the shower increases until the energy of the secondary particles is degraded to the
level where ionization losses dominate. At this point --as the particles only lose energy, are absorbed, or decay--
the density of particles starts to decline. 
In each generation about 20\% of the energy is transferred to the electromagnetic cascade that, 
ultimately, dissipates roughly 90\% of the primary particle's energy through ionization of the atmosphere atoms.
Dedicated observatories exist, such as the Pierre Auger Observatory~\cite{ThePierreAuger:2015rma} and 
the Telescope Array (TA)~\cite{AbuZayyad:2012ru}, 
that determine the energy and mass of the incoming UHECR in arrays of detectors by (i) sampling 
the fraction of the EAS particles that reach ground and/or, in moonless nights, by measuring 
(ii) the fluorescence photons that are produced by nitrogen molecules excited by the particle shower.\\

Usually the distance along the EAS axis is measured as a column density X (longitudinal shower profile, measured in g\,cm$^{-2}$),
indicating the amount of air traversed downstream from the top of the atmosphere in the direction of the shower propagation. 
For reference, the total vertical column density at sea-level is about 1\,000~g\,cm$^{-2}$, and the total column 
density for a shower traversing the atmosphere at zenith angle of $\theta = 60^\circ$ is twice larger.
A {\it vertically-incident} 10$^{20}$~eV proton produces about 10$^{11}$ secondaries at sea-level with energies above 90~keV in
the annular region extending from 8~m to 8~km off the shower axis. Of these, 90\% are $\gamma$'s, 9\% $e^\pm$, and 1\% $\mu^\pm$
plus charged hadrons. The mean energy of \elm\ particles is around 10 MeV and they transport 85\% of the total energy at ground level. 
These numbers change dramatically for the case of very inclined showers. For a primary zenith angle, $\theta > 60^\circ$, 
the electromagnetic component becomes attenuated exponentially with atmospheric depth, being almost completely absorbed 
at ground. For this reason, inclined showers with $\theta \approx 60^\circ$ are particularly useful to probe the hadronic 
and muonic properties of the cascades, and will be studied in detail in this work.\\

The determination of the original UHECR energy and mass is based on a detailed comparison of the 
EAS properties to the predictions of Monte Carlo (MC) models of the hadronic and electromagnetic 
development of the particle shower. 
Key EAS observables are the average depth of the shower maximum $\Xmax$ and the RMS width of its fluctuations 
$\smax$, the number and total energy of electrons ($\rm N_e, E_e$) and muons ($\rm N_\mu, E_\mu$) on ground for 
various shower zenith angle ($\theta$) inclinations.
The primary mass composition and energy of UHECR is thereby extracted by comparing the
experimental measurements to the results of full MC simulations of the EAS development in the atmosphere 
for different incoming species (protons, He, N, and Fe ions, mostly) at various initial candidate energies. 
This is commonly done with transport programs such as \corsika~\cite{Heck:1998vt}
interfaced to a set of hadronic interaction models such as \epos~\cite{Werner:2005jf}, \eposlhc~\cite{Pierog:2013ria},
\qgsjet~01~\cite{Kalmykov:1993qe,Kalmykov:1997te}, \qgsjetII~\cite{Ostapchenko:2010vb}, or
\sibyll~2.1~\cite{Ahn:2009wx} for the hadronic interactions, plus {\sc egs}4~\cite{Nelson:1985ec} for the \elm\ 
cascade evolution. All these hadronic interaction models 
describe the inclusive production of particles in high-energy proton and nucleus collisions based 
on basic quantum field-theory principles --such as unitarity and analyticity of scattering amplitudes--
as implemented in the framework of Gribov's Reggeon Field Theory (RFT)~\cite{Gribov:1968fc}, extended 
to take into account perturbative quantum chromodynamics (pQCD) scatterings in (multiple) harder 
parton-parton collisions via ``cut (hard) Pomerons'' (understood diagrammatically as a ladder of gluons). 
With model parameters tuned to reproduce the existing accelerator and collider results\footnote{The nominal LHC 
proton-proton center-of-mass (\cm) energy, $\sqrts$~=~14~TeV, corresponds to UHECR of $\ECR\approx 10^{17}$~eV colliding with air nuclei
at rest.}~\cite{dEnterria:2011twh}, all those MC generators are able to describe the overall EAS properties, although 
some ``anomalies'' persist in the UHECR data that cannot be easily accommodated. On the one hand, 
the $\Xmax$ and $\smax$ dependence on $\ECR$ indicates a change of cosmic ray composition from proton-dominated 
to a proton-iron mix above $\ECR \approx 10^{19}$~eV~\cite{Aab:2014kda,Abbasi:2014sfa},
but quantitative differences in $\Xmax$ exist, of up to 40~g\,cm$^{-2}$ among model predictions, that are not-fully understood, 
even though independently each MC event generator reproduces the  LHC data~\cite{Ostapchenko:2014mna,Pierog:2015epa,Ostapchenko:2016wtv}. 
On the other hand, in the same range of $\ECR$ energies, recent Auger measurements indicate about 30--60\% more 
muons on ground than expected by any of the MC models~\cite{Aab:2014pza,Aab:2016hkv}. 
A similar result was observed longer ago by the HiRes-MIA hybrid array, with a higher density of muons at 600~m 
from the shower's trajectory than expected from the (then current) hadronic interaction event generators~\cite{AbuZayyad:1999xa}. 
Those findings suggest that the best models of hadronic 
interactions are missing some physics ingredient. Either they do not account for processes that produce harder muons,
such as from \eg\ hard jets or heavy-quark (in particular charm~\cite{Engel:2015dxa}) decays, and/or they do not feed enough energy into 
the hadronic component of the EAS (such as \eg through an increased production of baryon-antibaryon pairs~\cite{Pierog:2006qv}). 
More speculative explanations have been put forward based on changes in the physics of strong interactions 
at energies beyond those tested at the LHC~\cite{Allen:2013hfa,Aloisio:2017ooo}.\\

The main purpose of this work is to test whether the aforementioned UHECR data--model differences can be 
explained as due to missing perturbative processes in the RFT-based approaches. For this purpose, hadronic collisions
are generated with the standard MC event generator used in particle physics, \pythia~6~\cite{Sjostrand:2006za} 
based on a purely pQCD framework that includes the production of rarer processes such as (multiple) 
high transverse momentum jets and heavy-quarks (charm and bottom). 
Despite its success in reproducing a large amount of experimental proton-(anti)proton collider data through tuned settings 
of its model parameters~\cite{Skands:2010ak}, \pythia\ has never been used to study cosmic rays interactions
as it cannot deal (so far) with the proton-nucleus or nucleus-nucleus interactions encountered 
in UHECR collisions with air nuclei. In order to overcome this limitation, we construct first 
a hydrogen atmosphere where the only target ``nuclei'' present are protons, with a density matching that of air,
using the fast \conex\ transport simulation~\cite{Bergmann:2006yz}.
Second, we interface \pythia~6, with various settings of its MC parameters, 
as well as the rest of RFT-based models, with \conex\ in order to generate the corresponding extended 
``air'' showers from incoming proton cosmic rays with energies $\ECR\approx~10^{14}$--$10^{20}$~eV,
and compare the properties of the resulting electromagnetic, hadronic, and muonic EAS components.



\section{Theoretical setup}

\subsection{\conex\ air-shower simulation}

For primary cosmic rays of energies above $\ECR\approx 10^{18}~eV$, full simulations of their EAS 
development performed with the \corsika\ program~\cite{Heck:1998vt} are time-consuming, and a systematic 
study of the shower properties for many variations of the underlying hadronic interaction models settings is prohibitive
in practical terms. A viable alternative consists in using a hybrid air-shower scheme, such as that implemented in 
\conex~\cite{Bergmann:2006yz}, effectively combining two main stages: an explicit MC simulation of particle cascading at energies 
above some chosen threshold $\rm E_{thr}$ (typically a factor of 100 smaller than the energy $\ECR$ of the incoming primary particle),
plus a numerical solution of the hadronic-electromagnetic cascade equations for sub-cascades at smaller energies. 
For the first part of the EAS development, the simulation of all particle interactions and decays is taken care of by 
the chosen interfaced high-energy hadronic interaction model (here \pythia~6, \eposlhc, \qgsjetII, \qgsjet~01, and \sibyll~2.1),
and the characteristics of all produced particles (type, energy, and slant depth position)
below $\rm E_{thr}$ are written into corresponding stacks or ``source terms''. Such sources 
provide the initial conditions for the subsequent fast numerical solution of the equations describing
the second part of the cascade, calculated only along the direction of the shower axis.
The final results are discretized energy spectra of all particles of different types at various depth shower positions.
In the first MC cascade step, \conex\ follows the propagation, interaction and decay (where applicable) of 
(anti)nucleons and (anti)hyperons, charged and neutral pions/kaons, whereas all other types of hadrons 
produced in interactions (including D and B heavy-quark hadrons in the case of \pythia) are assumed to decay immediately.
The MC \elm\ cascade is realized with the {\sc egs}4 code, complemented with the Landau-Pomeranchuk-Migdal effect 
for ultra-high energy $e^\pm$ and $\gamma$. In the second analytical step, the system of coupled \elm\ cascade
equations is based on the same interaction processes implemented in the MC (Bethe-Heitler for 
bremsstrahlung and pair production, Klein-Nishina for the Compton process, M\"oller and Bhabha processes,
as well as $\epem$ annihilation).
In order to generate the lateral distribution of the EAS at ground, low energy particles can be sampled from 
the energy distribution of particles along the shower axis produced by the cascade equations following the \seneca\ 
model approach~\cite{Drescher:2002cr}. Typically, hadrons below 300~GeV, muons at all energies and \elm\ particles 
with less than 10~GeV are then tracked again in the MC cascade, where Coulomb scattering and transverse momentum of 
the particles can be taken into account. Spatial and temporal distribution of particles are then similar to what 
can be obtained from a full MC cascade such as \corsika~\cite{conex3D,Drescher:2002cr}.
In this work, the \conex\ programme is initialized with a modified model of the earth atmosphere, 
changing N and O atomic nuclei by hydrogen alone with a density matching that of air, and interfaced 
with \pythia~6 as well as with the other four hadronic MC models. With such a setup we generate multiple 
``fixed-target'' proton-proton (pp) collisions over the range of energies $\ECR\approx~10^{14}$--$10^{20}$~eV.
Hereafter, the results shown for ``sea-level'' will refer to a ground at the sea level of 0~m ($\sim$1\,000~g\,cm$^{-2}$ 
vertical depth or $\sim$2\,000~g\,cm$^{-2}$ slant depth for shower with a zenith angle $\theta$ of 60$^\circ$).


\subsection{\pythia~6 Monte Carlo settings}

The basic ingredients of \pythia~6.428 are leading-order pQCD $2\to 2$ matrix elements, 
complemented with initial- and final-state parton radiation (ISR and FSR), convolved with 
parton distribution functions (PDFs) for the initial state, and the Lund string model~\cite{Andersson:1983ia} 
to describe the final parton hadronization. The infrared $1/\pT^4$ divergence of the hard 
(multi)parton cross section, when the transverse momentum of the minijet $\pT\to 0$, is regularized by a cutoff $Q_0$, such that $1/\pT^4 \to 1/(\pT^2+Q_0^2)^2$,
that depends on a power $\epsilon$ of the pp \cm~energy: 
$Q_0^2(s)=Q_0^2(s_0)\cdot(s/s_0)^\epsilon$, where $Q_0(s_0)\approx$~2.5~GeV is a reference value 
at a given \cm energy $\sqrt{s_0}\approx 7$~TeV. 
The values of the $Q_0$ and $\epsilon$ parameters impact the total hadron multiplicity in a given pp collision:
a {\it higher} scaling power of the infrared cut-off implies a {\it slower} increase of the overall hadronic activity.
Other non-perturbative ingredients of \pythia\ include a Regge-based modeling of diffractive 
processes~\cite{Schuler:1993wr}, plus a model for the underlying-event issuing from 
multi-parton interactions (MPI), soft scatterings, and beam-remnants~\cite{Sjostrand:1987su}.
The MPI are treated perturbatively based on 
an impact-parameter-dependent transverse overlap of the colliding protons, described by a Gaussian profile
in all the settings considered here.\\

In this work, seven different sets of model parameters (tunings) of \pythia~6.428 are considered
via the \ttt{PYTUNES} switch for the description of semihard (ISR and FSR showering) 
and non-perturbative (hadronization) dynamics. The default settings are those of the  central ``Perugia'' (350) 
tune fitted to underlying event (UE), minimum bias (MB), and Drell-Yan  measurements from 2011 pp 
collisions at the LHC~\cite{Skands:2010ak}, using the CTEQ5L PDFs~\cite{Lai:1999wy}. The production and decay 
of secondary charm and bottom hadrons is handled directly by \pythia~6, namely we do not consider their possible 
direct interaction with the ``target'' partons of atmospheric protons. We also run the 350 tune with  
heavy-quark production explicitly switched-off\footnote{Technically, this is done by setting 3 flavours only ($u, d, s$) in 
ISR (\ttt{MSTP(58)=3}) and FSR (\ttt{MSTJ(45)=3}), and by switching off gluon splittings into charm and bottom: 
\ttt{IDCc=MDCY(21,2)-1+4},\ttt{IDCb=MDCY(21,2)-1+5}, \ttt{MDME(IDCc,1)=0}, \ttt{MDME(IDCb,1)=0}.}
in order to estimate the impact of charm and bottom production on the shower properties.
In addition, the following five other tune 
variations, based on 2012 LHC data using the CTEQ6L parton densities~\cite{Pumplin:2002vw}, with slightly 
increased strangeness ($s\bar{s},\eta,\eta'$) production and softer baryon spectra compared to the 350 tune, are used:
\begin{description}
  \setlength\itemsep{-0.2em}
\item (i) tune 371 with high ISR and FSR obtained evaluating the QCD coupling at a scale $\alphas(\pT/2)$, 
\item (ii) tune 372 with low ISR and FSR obtained using  $\alphas(2\pT)$, 
\item (iii) tune 380, using only gluon-gluon processes at low-$\pT$, without valence-quark scattering (\ttt{PARP(87)=0}),
\item (iv) tune 381, with lower values of the $Q_0$ and $\epsilon$ parameters leading to a higher amount of UE activity,
\item (v) tune 382, with higher values of the $Q_0$ and $\epsilon$ parameters leading to a lower amount of UE activity. 
\end{description}
Table~\ref{tab:pythiaTunes} provides a summary of the seven \pythia~6 MC tune settings considered in this work.
We note also that the overall properties of particle production of the chosen set of \pythia~6 tunes are very similar 
to those obtained with the latest tunes of the \pythia~8 version of the code~\cite{Sjostrand:2007gs}, as discussed in Ref.~\cite{dEnterria:2016oxo}.
\begin{table}[htbp]
\begin{center}
{\footnotesize
\begin{tabular}{lcccccc}\hline
\pythia~6.428 Perugia tune & PDF     & $Q_0$ cutoff at      & $Q_0$ scaling    & ISR/FSR scale       & Hadronization  \\ 
 \ttt{PYTUNES} number (main features)   &         & $\sqrt{s_0}$~=~7 TeV & power $\epsilon$ & $\alphas(k\cdot\pT)$  &  \\\hline
 350 (central tune 2011)        & CTEQ5L1 & 2.93 GeV             & 0.265            & $k=1$  & $s\bar{s},\eta,\eta'$ suppr.\,=\,95,63,12\%  \\ 
 350, noHQ (central 2011; no c-,b-quarks) & CTEQ5L1 & 2.93 GeV& 0.265   & $k=1$  & $s\bar{s},\eta,\eta'$ suppr.\,=\,95,63,12\% \\ 
 371 (var. 2012, high rad.)& CTEQ6L1 & 2.72 GeV & 0.25 & $k=1/2$&$s\bar{s},\eta,\eta'$ suppr.\,=\,92,70,13.5\%; softer baryons \\ 
 372 (var. 2012, low rad.) & CTEQ6L1 & 2.60 GeV & 0.23 & $k=2$  &$s\bar{s},\eta,\eta'$ suppr.\,=\,92,70,13.5\%; softer baryons \\ 
 380 (var. 2012, $gg$ only at low-$\pT$)& CTEQ6L1 & 2.65 GeV & 0.245& $k=1$  &$s\bar{s},\eta,\eta'$ suppr.\,=\,92,70,13.5\%; softer baryons  \\ 
 381 (var. 2012, higher UE)& CTEQ6L1 & 2.46 GeV & 0.23 & $k=1$  &$s\bar{s},\eta,\eta'$ suppr.\,=\,92,70,13.5\%; softer baryons \\ 
 382 (var. 2012, lower UE) & CTEQ6L1 & 2.92 GeV & 0.26 & $k=1$  &$s\bar{s},\eta,\eta'$ suppr.\,=\,92,70,13.5\%; softer baryons\\ 
\hline
\end{tabular}
}
\caption{Details of the various ingredients controlling the  semi-hard and non-perturbative dynamics in
 the seven tunes of \pythia~6.428 used in this work. See text and~\cite{Skands:2010ak} for details.}
\label{tab:pythiaTunes}
\end{center}
\end{table}

\subsection{RFT-based Monte Carlo event generators}

In the standard RFT-based MCs used in cosmic-ray physics (\eposlhc, \qgsjetII, \qgsjet~01, and \sibyll~2.1),
hadronic collisions are generated starting from a construction of the scattering amplitude to determine the total and elastic cross sections 
with intercept and slope of the Pomeron ($\Pom$)  Regge trajectory, Pomeron-hadron couplings, \etc\ fixed from experimental data.
Inelastic events are generated ``cutting'' diagrams involving multiple-$\Pom$ exchanges 
based on the so-called Abramovskii-Gribov-Kancheli rules~\cite{Abramovsky:1973fm}.
Cut Pomerons correspond to color flux tubes, treated as strings extended between the colliding partons, 
that subsequently fragment into hadrons separately following various 
hadronization models with parameters fitted to reproduce the data. Leading-order 
pQCD scatterings are modeled, above a scale $Q_0$, through multiple ``cut hard Pomerons'' diagrams. 
Since fixed-target collisions of protons with 10$^{20}$~eV energies involve gluon interactions with fractional momenta 
$x \approx \pT/\sqrts \approx$~10$^{-7}$, three orders-of-magnitude smaller than those currently probed in PDF extractions,
the models include various approaches to deal with the onset of non-linear
(gluon fusion) effects saturating the growth of the PDFs as $x\to$~0. 
The different generators used here differ in various approximations for the collision configurations 
(\eg\ for the number of cut-$\Pom$ and the energy-momentum partition among them), 
the treatment of diffractive and perturbative contributions as well as of low-$x$ PDFs, 
and the details of particle production from string fragmentation.
In the \qgsjet\ models, the transverse profile of the proton (the underlying PDF in the impact-parameter direction)
is effectively Gaussian, whereas in \sibyll~2.1 it is taken as the Fourier transform of the proton electric form factor, 
resulting in an energy-independent exponential fall-off of the transverse PDF. The harder form of the \sibyll\
form factor allows a greater retention of energy by the leading particle, and hence less of it available for the ensuing shower.
Table~\ref{tab:MCs}  provides a summary of the particle-production properties of the four RFT-based MC event generators 
considered in this work.

\begin{table}[htbp]
\begin{center}
\begin{tabular}{lccc}\hline
Model (version)   &  $Q_0$ hard-soft threshold & Low-$x$ PDF                      & Hadronization \\\hline
\sibyll~2.1       & 1.0 GeV+ $\exp{\sqrt{\ln{s}}}$      & GRV~\cite{Gluck:1998xa}        & Lund string fragmentation\\
\qgsjet~01        & 2.0 GeV      & constant         & own string fragmentation \\
\qgsjetII         & 1.6 GeV      & non-linear $\Pom$ corrections + pQCD  & own string fragmentation\\
\eposlhc          & 2.0 GeV      & parametrized soft + pQCD       & area-law string fragmentation + collective flow \\\hline
\end{tabular}
\caption{Details of the main ingredients controlling the non-perturbative and semi-hard dynamics
present in the RFT-based event generators used in this work.}
\label{tab:MCs}
\end{center}
\end{table}

\section{Results}

The properties of the extended atmospheric showers predicted by the pQCD- and RFT-based
MC generators are studied by running the simulation setup described in the previous Section. 
For each one of the 14 incoming cosmic-ray proton energies over the range 
$\ECR\approx~10^{14},10^{14.5}\cdots 10^{20},10^{20.5}$~eV, one thousand air-showers are generated using \conex\ interfaced 
with the seven tunes of \pythia~6, as well as with \eposlhc, \qgsjetII, \qgsjet~01, and \sibyll~2.1 (with their default settings), 
totaling 154\,000 EAS generated and statistically studied. The generic features of the generated showers,
given by $\Xmax$ and $\smax$, as well as the properties of their electromagnetic, hadronic, and muonic 
components are discussed in detail in the following subsections.

\subsection{Generic properties of high-energy pp collisions}

The relationship between the key ingredients of hadronic interaction MCs and EAS observables can be extracted 
using a generalized Heitler type model~\cite{Matthews:2005sd} and
has been numerically studied  in detail in~\cite{Ulrich:2010rg}. 
The average depth $\Xmax$ and fluctuations $\smax$ of the shower maximum  depend chiefly on the characteristics 
of multiparticle production of the first few generations of hadronic interactions in the shower.
The key model ingredients are (i) the {\it inelastic pp cross section} $\sigma_{\rm inel}$ (combined with a Glauber model
to derive the proton-air cross section~\cite{Auger:2012wt}), 
(ii) the {\it multiplicity} ($N_{ch}$) of the primary and subsequent very high-energy interactions, which affects 
how the energy is distributed to secondary particles and corresponding subshowers, 
and (iii) the {\it inelasticity} $K = 1 - E_{\rm lead}/\ECR$ or fraction of the primary particle energy transferred 
to secondary particles after removing the most energetic ``leading''  hadron emitted at very forward rapidities. 
In addition, the {\it mean transverse momentum} of the particles produced, which is closely related to the peak of the 
(mini)jet production cross section around the scale $Q_{0}$, is a sensitive quantity of the modeling of the 
transition from soft to hard scatterings. The level of agreement between the MC predictions for many observables and the 
LHC data was thoroughly discussed in~\cite{dEnterria:2011twh,dEnterria:2016oxo}, where it was found that the models 
overall bracketed the experimental measurements and only small modifications, such as those that led to the 
\pythia~6 tunes of Table~\ref{tab:pythiaTunes} and to the new \eposlhc\ release, were required.

\begin{figure}[htbp!]
\centering
\includegraphics[width=0.49\textwidth]{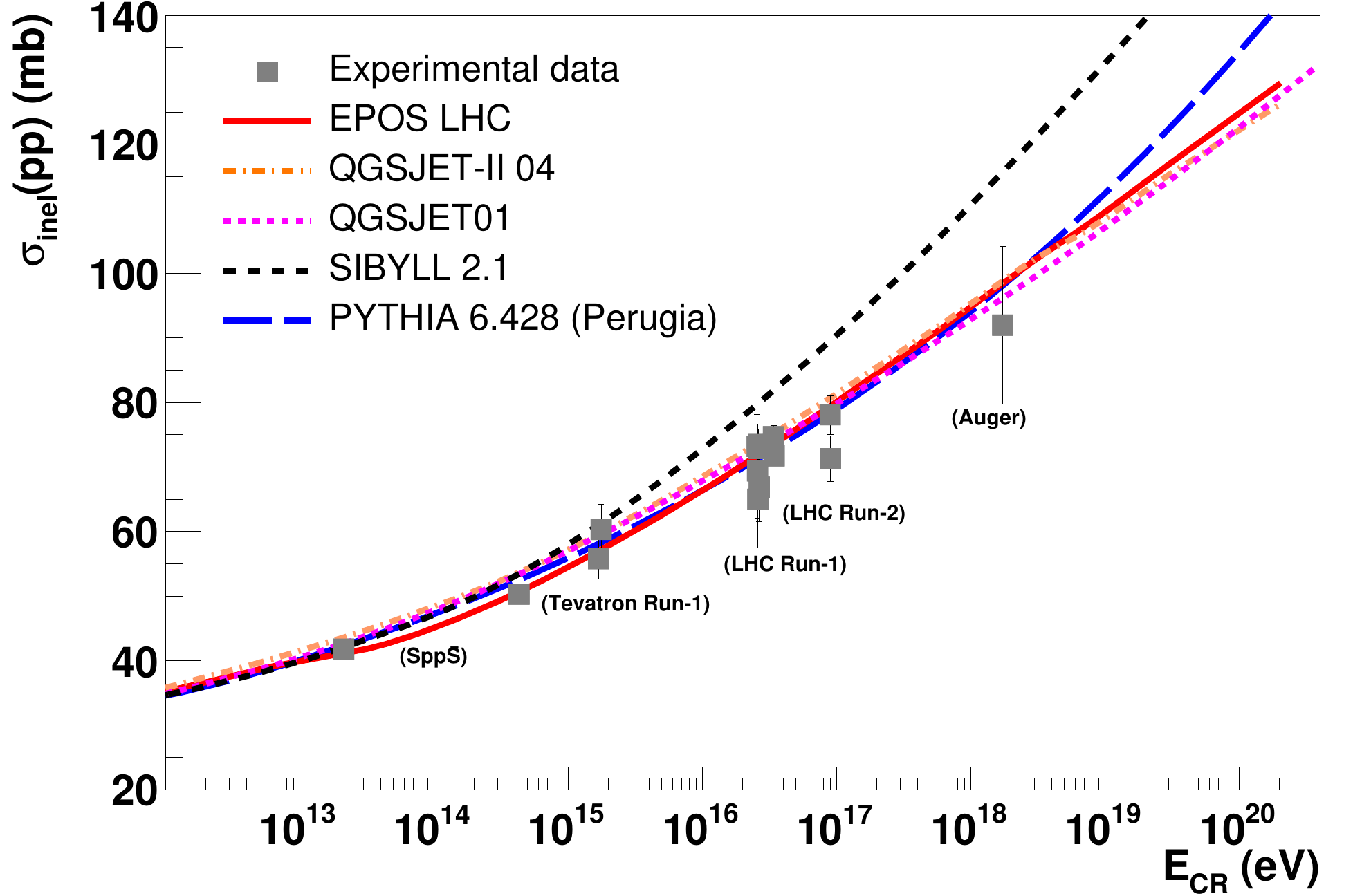}
\includegraphics[width=0.49\textwidth]{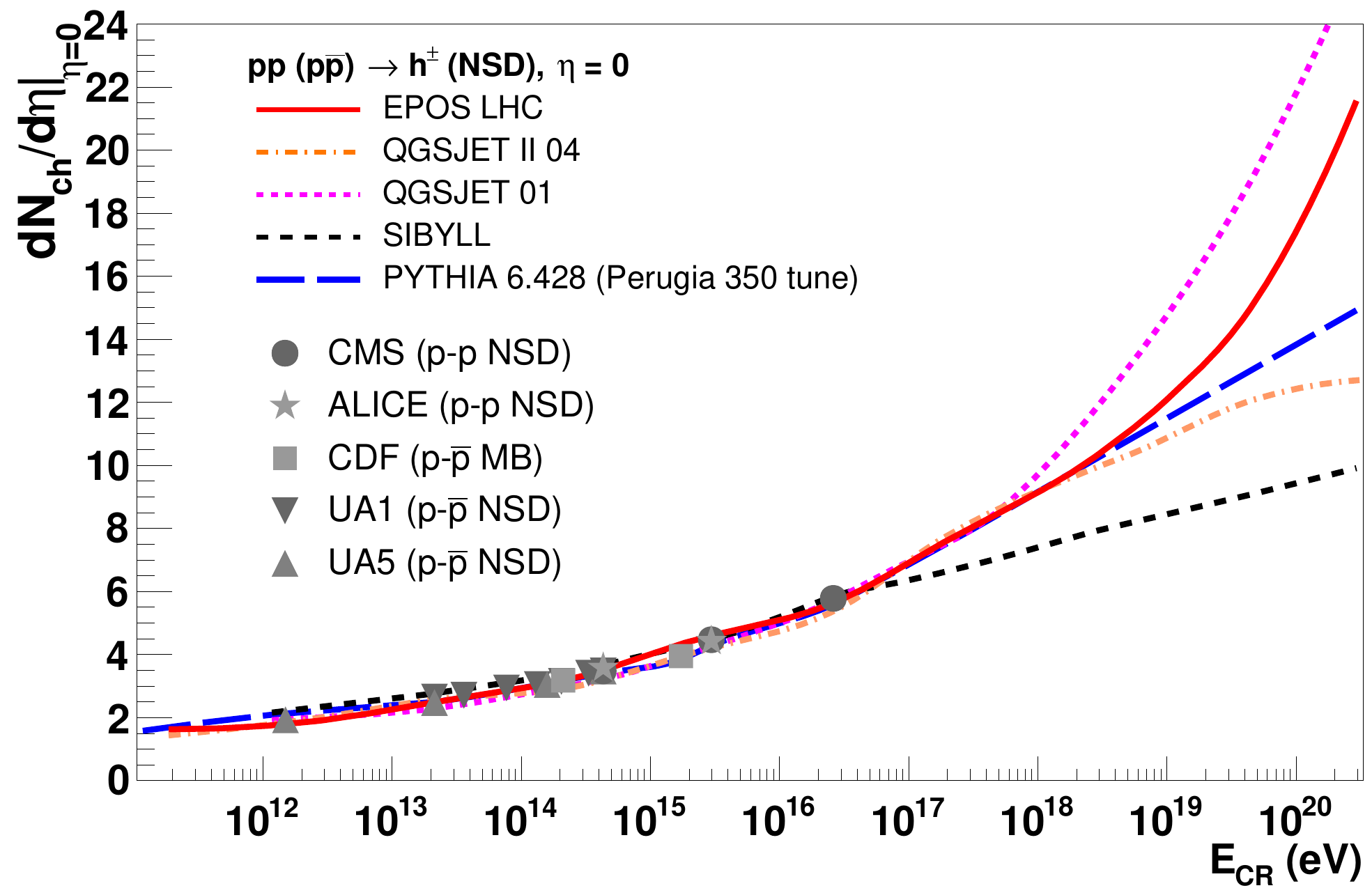}
\includegraphics[width=0.49\textwidth]{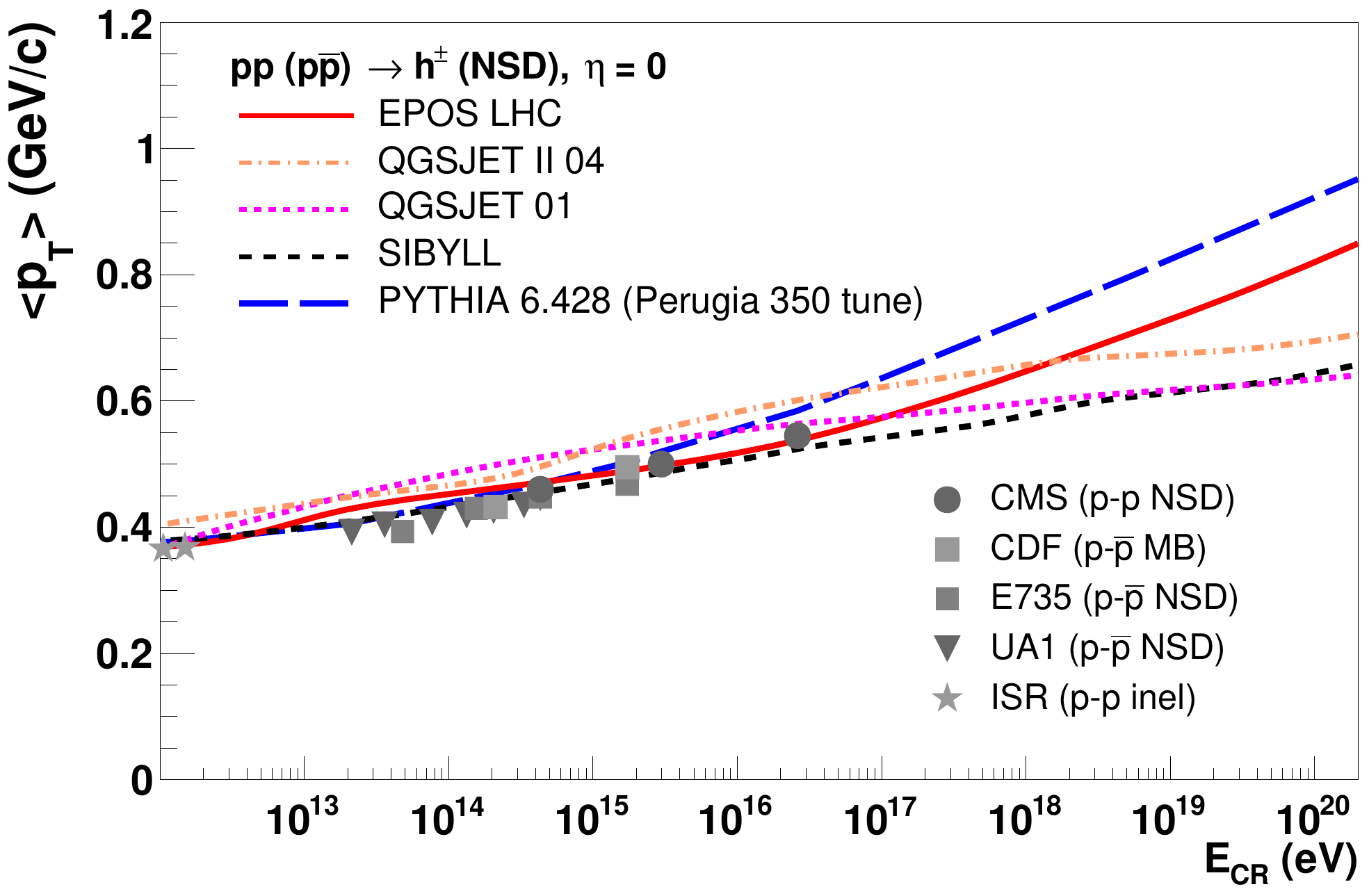}
\includegraphics[width=0.49\textwidth]{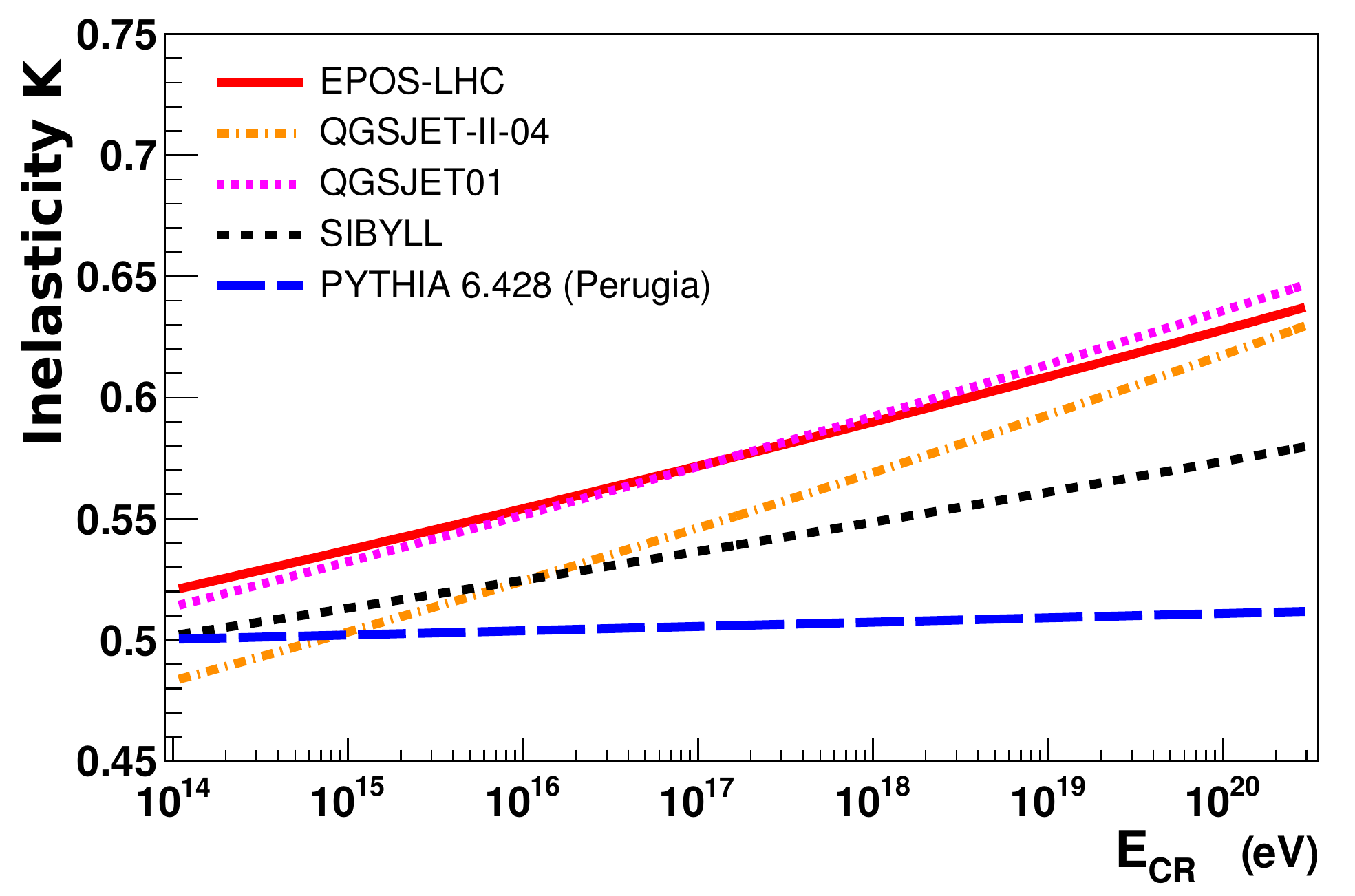}
\caption{Cosmic-ray energy dependence of key model ingredients of the MC event generators considered in this work: 
Inelastic pp cross section (top left), midrapidity 
charged particle multiplicity density (top right), mean transverse momentum (bottom left), and inelasticity 
(bottom right). Experimental data points, either from  non-single-diffractive (NSD) or ``minimum bias'' (MB)
collisions are from the compilations of Refs.~\cite{dEnterria:2011twh,dEnterria:2016oxo}.}
\label{fig:sigma_pp_vs_sqrts}
\end{figure}

Figure~\ref{fig:sigma_pp_vs_sqrts} shows the dependence on incoming cosmic-ray energy of the pp inelastic cross section (top left),
charged particle density at midrapidity (top right), mean transverse momentum (bottom left), and pp inelasticity 
(bottom right). For \pythia~6, only the default 350 tune is shown as other tunings, including that inhibiting heavy-flavor production, 
give identical or very similar results for such inclusive quantities. In the region where collider data exist, below $\ECR\approx 10^{17}$~eV, 
all MC models show a similar energy dependence consistent with the experimental results. Above $10^{17}$~eV, increasingly
bigger differences appear, with \pythia~6 having higher $\meanpt$ and smaller inelasticity than the rest of models, but
being in the ``average'' region with regards to charged particle multiplicity and inelastic cross section values. The largest model differences 
show up in the prediction of the energy dependence of the inelasticity, where \pythia~6 features an almost flat behavior 
with proton energy, to be compared to a 25--30\% increase observed for \eposlhc, \qgsjetII, and \qgsjet~01 between 10$^{14}$~eV 
and 10$^{20}$~eV. Since $\Xmax$ and its average fluctuation $\smax$ are mostly driven by the pp inelastic cross section 
and inelasticity, the EAS simulated with \pythia~6 feature larger penetration in the atmosphere than those from 
the rest of event generators, as discussed next. 

\subsection{Generic features of proton-induced EAS}

\begin{figure}[htbp!]
\centering
\includegraphics[width=0.49\textwidth]{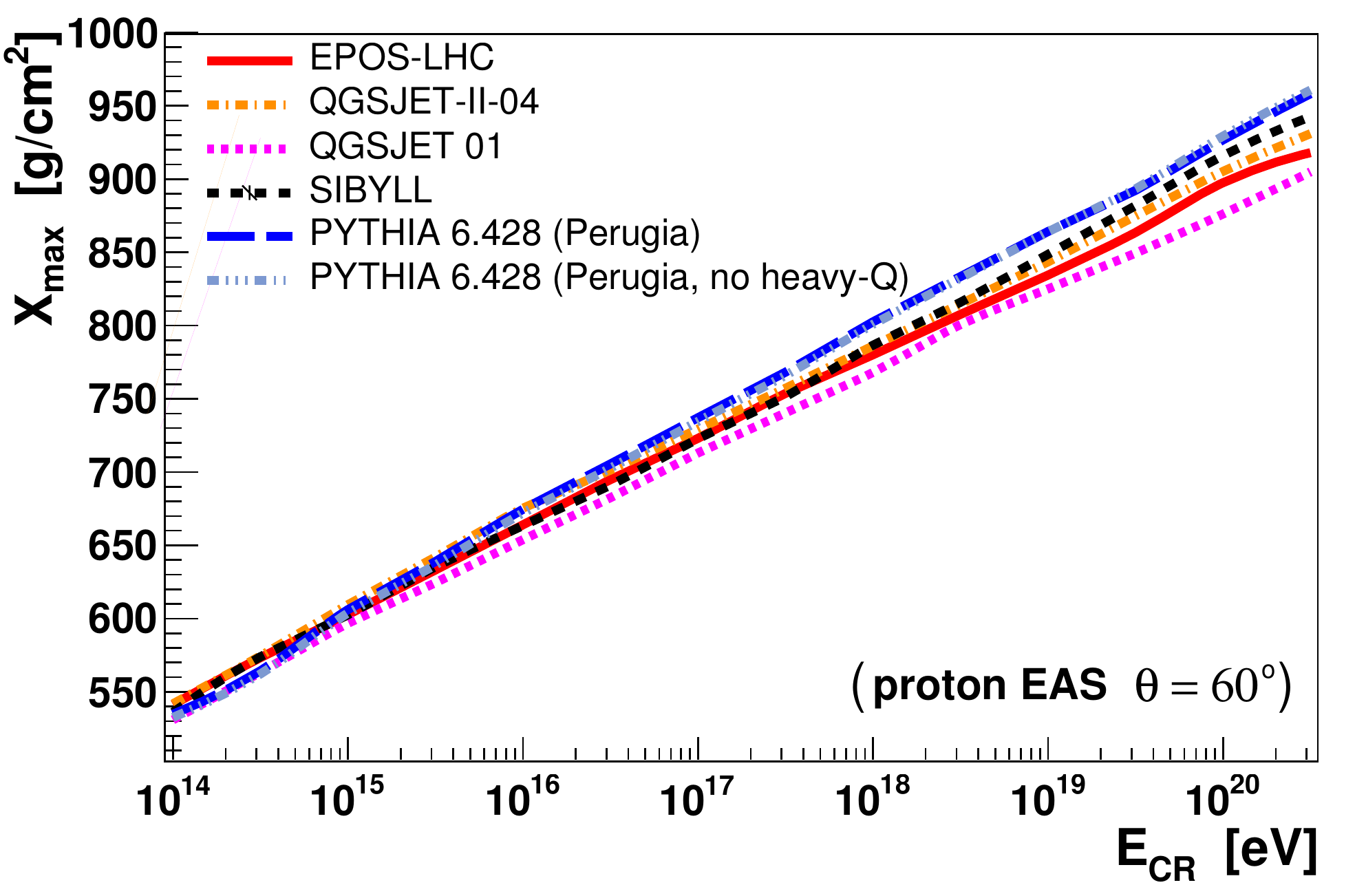}
\includegraphics[width=0.49\textwidth]{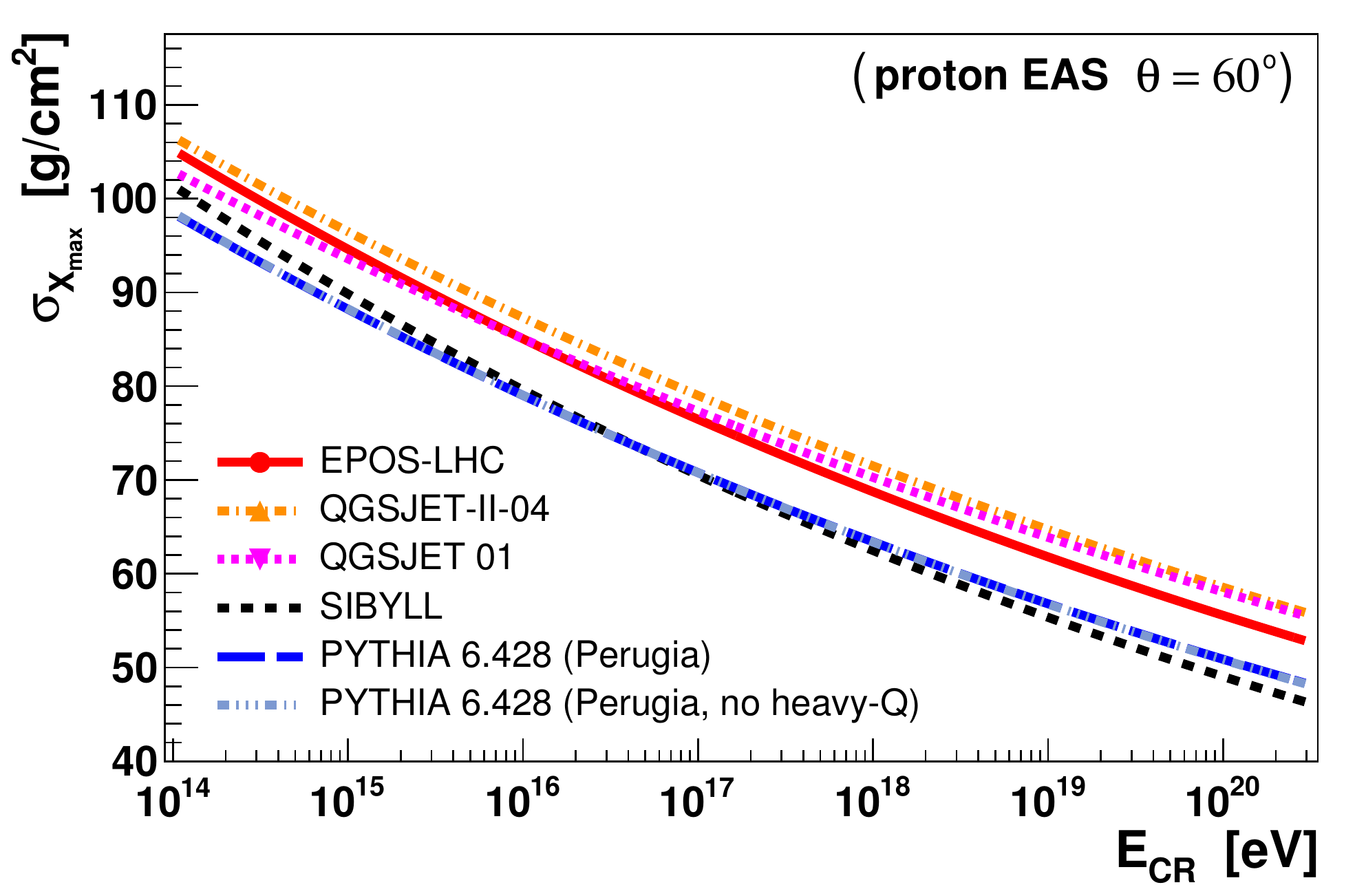}
\caption{Mean slant depth of the shower maximum $\Xmax$ (left) and width of its associated fluctuations $\smax$ (right) 
for inclined ($\theta = 60^\circ$) proton-induced showers as a function of the incoming cosmic-ray energy ($\ECR$), predicted by the 
six MC event generators considered here.}
\label{fig:Xmax_sigmaXmax_vs_E}
\end{figure}

Figure~\ref{fig:Xmax_sigmaXmax_vs_E} shows the average position of the shower maximum in the atmosphere $\Xmax$ (left) 
and the width of the fluctuations of the shower maximum position $\smax$ (right) as a function of the incident cosmic-ray proton energy for 
inclined showers ($\theta = 60^\circ$). 
As expected for showers generated with our ``Jupiter-like'' hydrogen atmosphere, the elongation rate is higher than for standard 
air showers~\cite{Pierog:2017nes}, leading to a high value of $\Xmax$ at high energy. 
Indeed the slope of the energy evolution of the inelastic cross section is about twice larger in the case of 
\pp\ compared to p-Air interactions~\cite{Pierog:2017nes}. The cross section itself is about three times smaller in \pp\ 
than in p-Air, leading to larger values of $\smax$ at low energy. 
Although all MCs predict relatively similar values for both quantities, 
\pythia~6 (all tunes, indistinguishably) features the largest $\Xmax$ values, \ie\ the largest penetration in the 
atmosphere, and the lowest $\smax$, \ie\ the smallest fluctuations in the altitude where the shower maximum appears.\\

In general, the model that gives closest (resp. farthest) results to \pythia~6\ is \sibyll~2.1 (resp. \qgsjet~01),
whereas \eposlhc\ shows an intermediate behavior among all models. 
At $\ECR\approx10^{20}$~eV, \pythia~6 predicts a shower maximum at a depth that is $\sim$80 (50)~g\,cm$^{-2}$ 
larger than that of \qgsjet~01 (\eposlhc). More surprising are the increasing differences among RFT model predictions. 
The difference in $\Xmax$ between \eposlhc\ and \qgsjetII\ is reduced from about 20~g\,cm$^{-2}$ for p-Air, to less 
than 10~g\,cm$^{-2}$ for \pp, the \qgsjetII\ values being larger than those from \eposlhc, a result reversed 
compared to that found for air showers~\cite{Pierog:2017nes}. This is a clear sign that nuclear effects play a non-negligible role, 
and are an important source of uncertainties, in the hadronic MC simulations.
In terms of the fluctuations of the altitude of the maximum shower, 
\pythia~6\ and \sibyll~2.1 share very similar behavior with about $5$--10~g\,cm$^{-2}$ smaller fluctuations than the 
rest of models, with \qgsjetII\ giving the largest values of $\smax$. The coincident results of \pythia~6 and \sibyll~2.1 
are largely accidental because, as can be understood from Fig.~\ref{fig:sigma_pp_vs_sqrts}, they are due to a similar net 
cancellation of many different underlying physical effects. Whereas the larger penetration of \pythia~6 showers is 
mostly due to a low pp inelasticity (Fig.~\ref{fig:sigma_pp_vs_sqrts}, bottom right), in the \sibyll~2.1 case this 
is due to the small number of particles produced per pp collision (Fig.~\ref{fig:sigma_pp_vs_sqrts}, top right).
The results of Fig.~\ref{fig:Xmax_sigmaXmax_vs_E} also show that switching-off heavy-quark production in \pythia~6 
has no effect on such inclusive EAS properties.

\subsection{Electromagnetic properties of proton-induced EAS}

\begin{figure}[htbp!]
\centering
\includegraphics[width=0.49\textwidth]{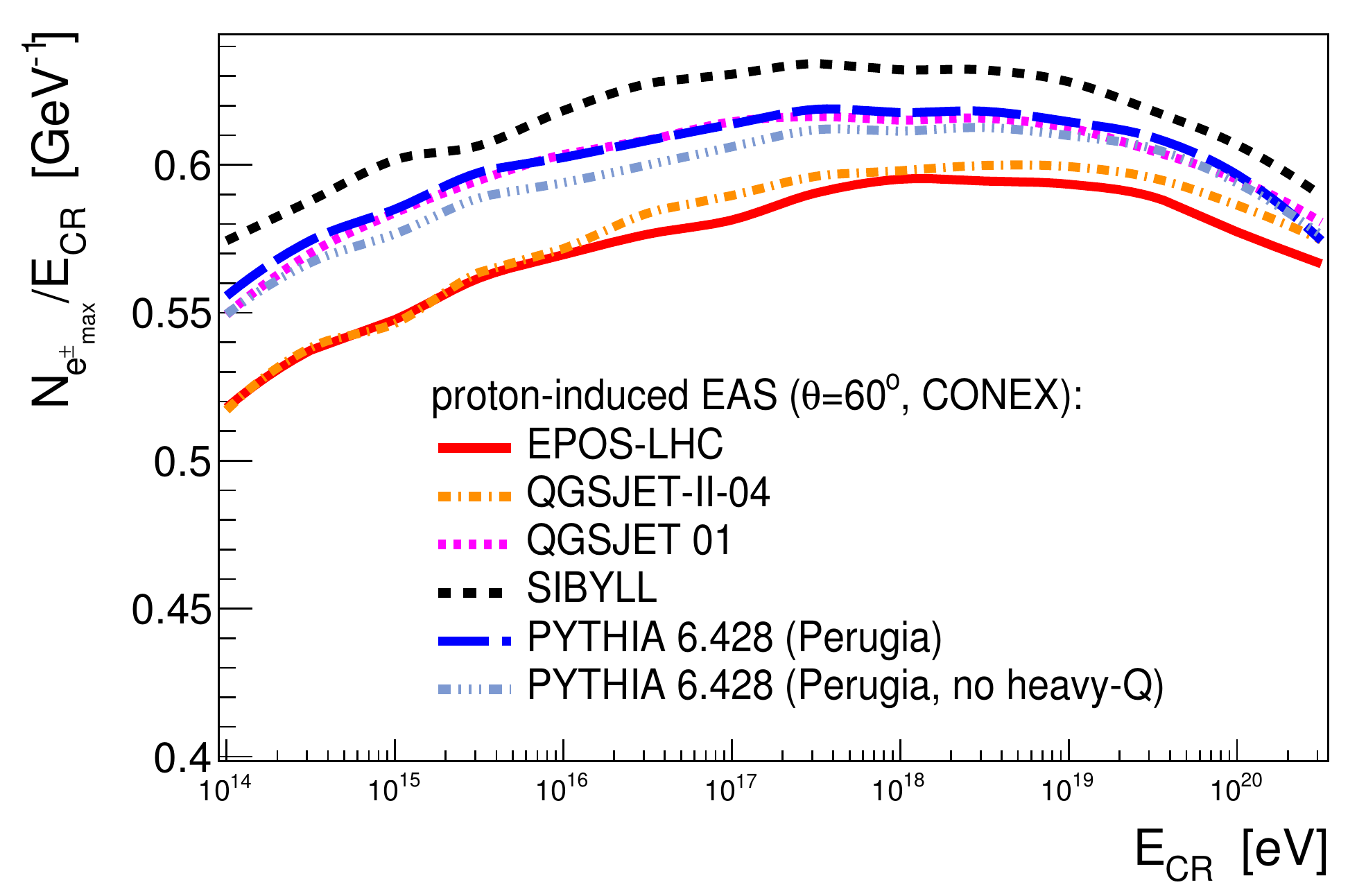}
\includegraphics[width=0.49\textwidth]{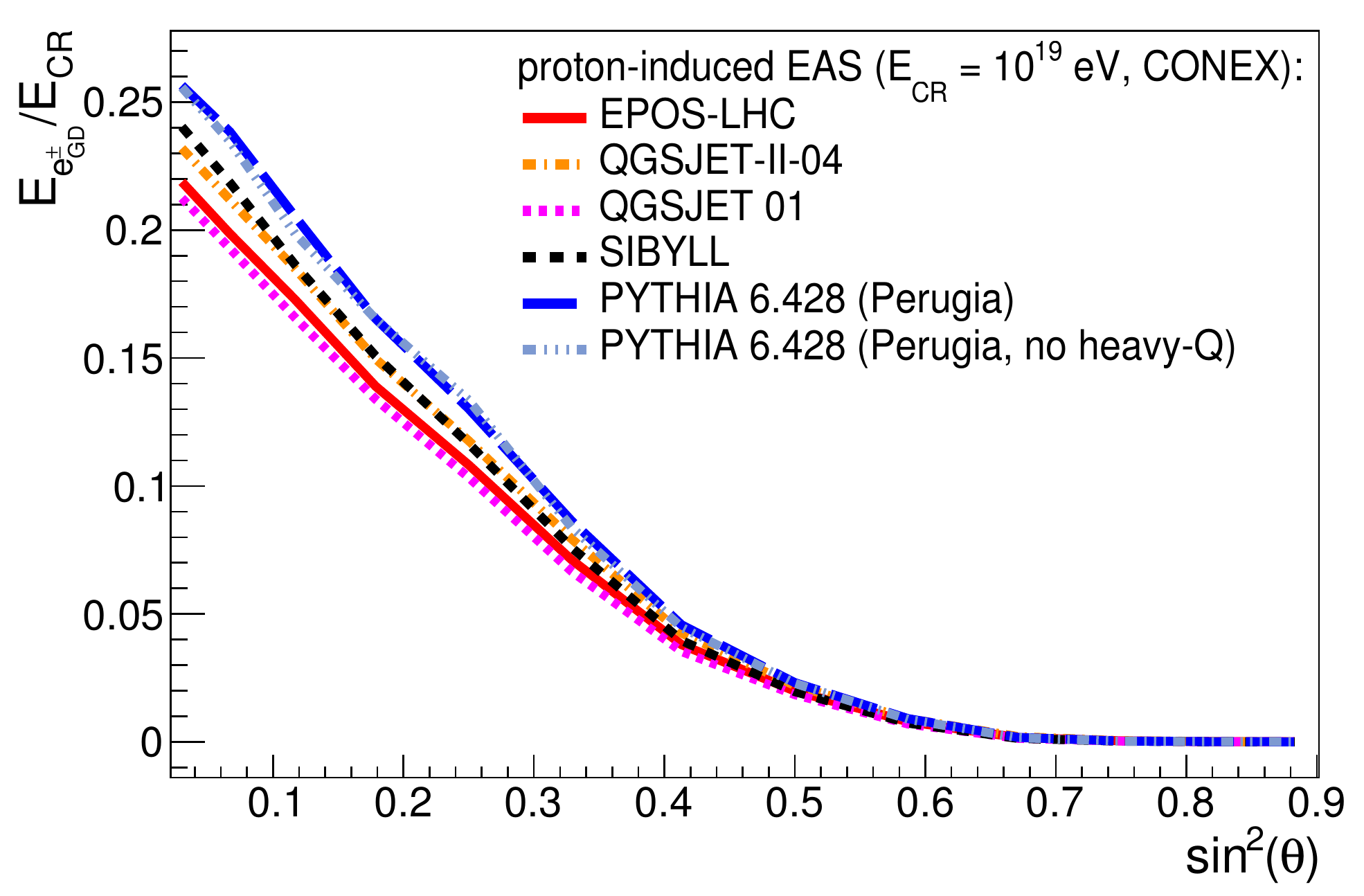}
\caption{Number of electrons and positrons at the shower maximum (normalized by the cosmic-ray energy $\ECR$) 
as a function of $\ECR$ for inclined proton-induced showers with $\theta = 60^\circ$ (left), and fraction of shower 
\elm\  energy at sea-level as a function of the (squared-sine) zenith angle for incoming protons with $\ECR = 10^{19}$~eV  
(right), predicted by the six MC event generators considered in this work.}  
\label{fig:Eem_vs_E}
\end{figure}

About two-thirds of the secondary particles produced in a hadronic collision are pions, with similar amounts of $\pi^0$,
$\pi^+$ and $\pi^-$ created. The neutral pions decay almost immediately into two photons which, at their turn, 
generate $e^+e^-$ pairs which then lose further energy through new $\gamma$ radiation, thereby generating 
an electromagnetic cascade in the atmosphere. Being well-known QED processes (bremsstrahlung, pair production, 
and ionization), the \elm\ part of the shower is well under control from a theoretical point of view. 
Figure~\ref{fig:Eem_vs_E} (left) shows the total number of electrons and positrons at the shower maximum for inclined 
showers ($\theta = 60^\circ$) as a function of cosmic-ray proton energy $\ECR$,  
for all MCs considered in this work. 
The distribution is normalized by the $\ECR$ value in order to display features that are otherwise difficult to discern in a steeply-falling spectrum.
All models predict a similar amount of $e^\pm$ at shower maximum (within $\sim$10\%), with \pythia~6 and \qgsjet~01 being
in-between \sibyll~2.1 and \eposlhc\ (that feature, respectively, the largest and lowest N$_{\rm e^{\pm} max}$ values). 
All models show a moderate increase of N$_{\rm e^{\pm} max}$ up to $\ECR \approx 10^{18}$~eV, followed by a decrease beyond that energy. 
This is due to the fact that above this energy, $\pi^0$ start to collide with the target nuclei, rather than decay, 
giving less energy to the \elm\ component of the shower.

The zenith-angle dependence of the fraction of UHECR energy carried by $e^\pm$ and $\gamma$ at ground
for proton-induced EAS with $\ECR = 10^{19}$~eV is shown in  
Fig.~\ref{fig:Eem_vs_E} (right). 
Inclined showers ($\theta = 60^\circ$) are on the right of the plot ($\sin^2\theta=0.75$), whereas
the leftmost values are for fully vertical EAS ($\sin^2\theta=0$). One sees that the amount of
electron energy at ground is $\sim$25\% for vertical showers, decreasing to zero with decreasing CR 
incident angle in the atmosphere. The more vertical the shower is, the less comparatively absorbed its \elm\ component is,
and the closer the ground is to $\Xmax$. Thus, here again, \pythia~6 features more 
\elm\ ground energy than the rest of the models for more vertical showers, simply due to its predicted 
comparatively larger EAS penetration.
We note also that switching-off heavy-quark production in \pythia~6 has no significant effect on the electromagnetic EAS properties.


\subsection{Hadronic properties of proton-induced EAS}

\begin{figure}[htbp!]
\centering
\includegraphics[width=0.49\textwidth]{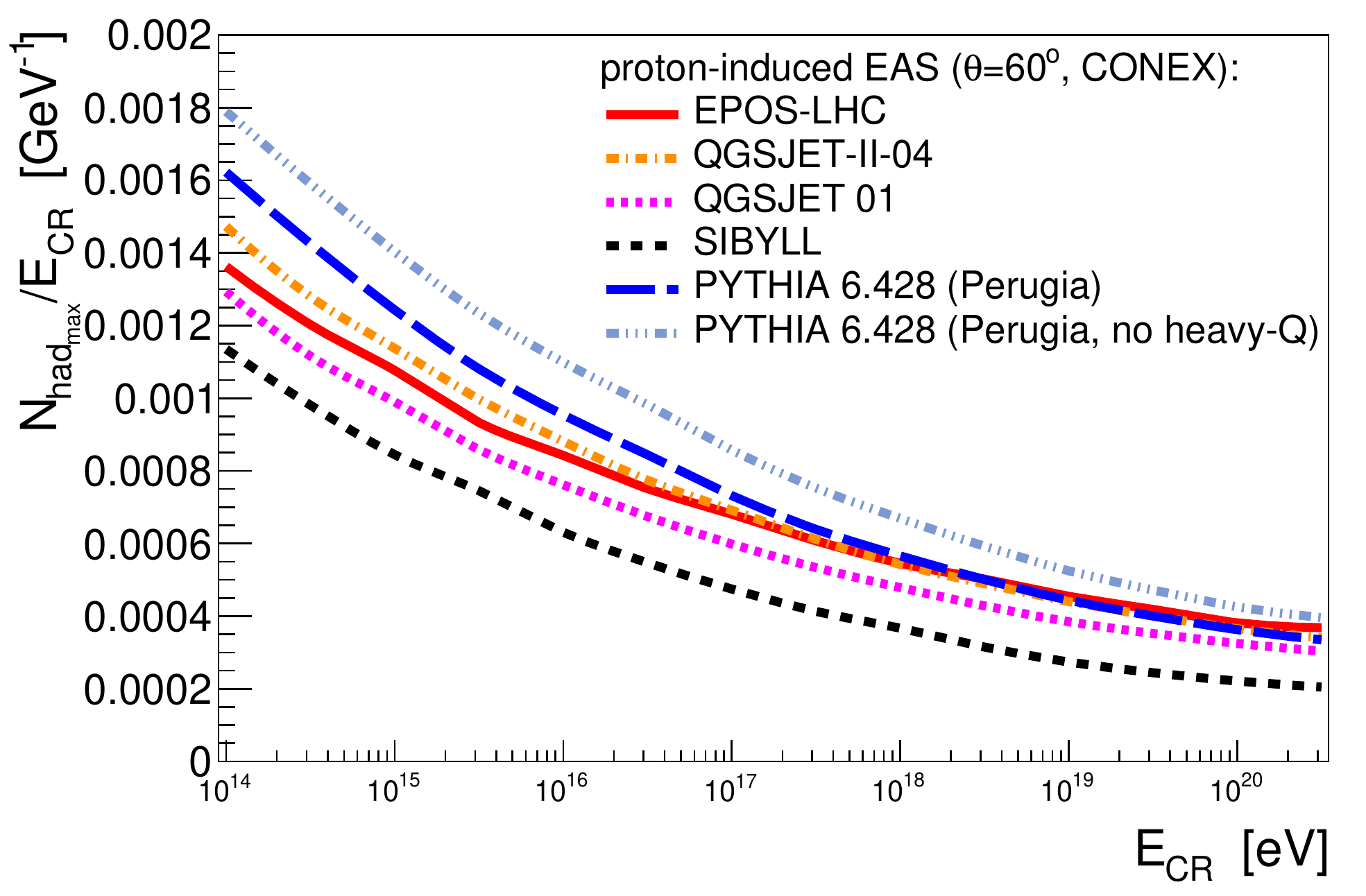}
\includegraphics[width=0.49\textwidth]{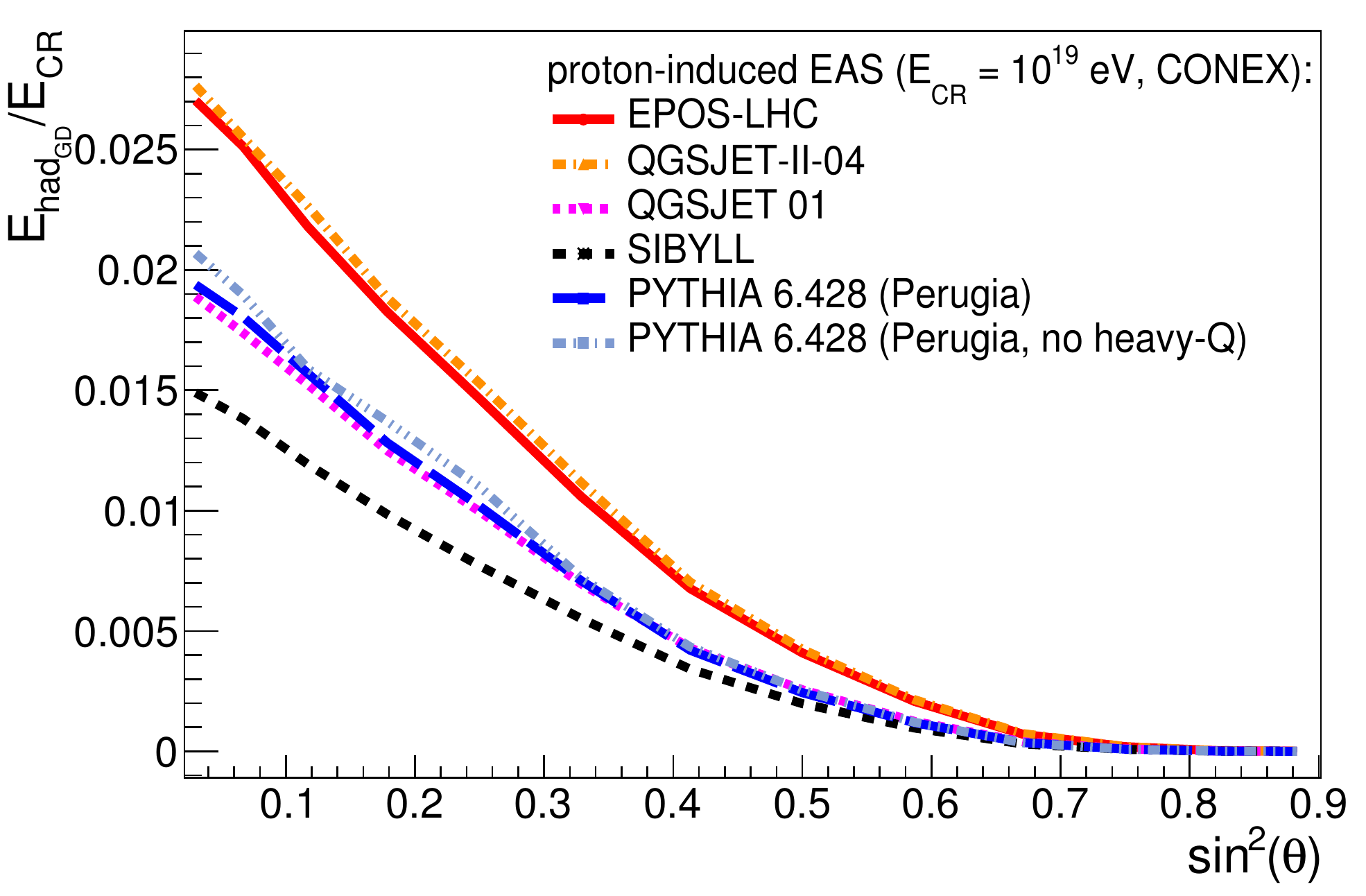}
\caption{Number of hadrons at shower maximum, normalized by the total CR energy, for inclined proton-induced showers 
($\theta = 60^\circ$) as a function of cosmic-ray energy $\ECR$ (left) and fraction of the total energy of the 
shower carried by hadrons at sea-level for proton-induced EAS with $\ECR\approx$~10$^{19}$~eV as a function of the 
(squared-sine) zenith angle of the incoming cosmic ray (right), predicted by the six MC event generators considered here.}  
\label{fig:Ehad_vs_E}
\end{figure}

At variance with the electromagnetic properties of the proton-induced showers, that are very similar among
MC models as shown in the previous section, larger differences appear in the hadronic properties of the EAS 
predicted by the different event generators.
Figure~\ref{fig:Ehad_vs_E} (left) shows the total number of hadrons produced at shower maximum (left) as a function 
of cosmic-ray proton energy $\ECR$ for inclined showers ($\theta = 60^\circ$) normalized by the $\ECR$ value.
(
Figure~\ref{fig:Ehad_vs_E} (right) shows the UHECR energy fraction carried by hadrons at ground for proton-induced EAS with 
$\ECR = 10^{19}$~eV as a function of the (squared-sine of the) zenith angle of the incoming cosmic ray (right) for all MCs 
considered in this work. It is interesting to notice that at variance with the increasing number of $e^\pm$ with $\ECR$ 
(Fig.~\ref{fig:Eem_vs_E} left), less and less hadrons are present at shower maximum for larger primary energies. This is so 
because the maximum  of the EAS is reached after an increasing number of hadronic generations, where about 20--30\% of 
the energy is given to the \elm\ component via $\pi^0$  decay.
In terms of the number of hadrons at shower maximum (left panel), \pythia~6 (in particular {\it without} charm and bottom production)
predicts the largest values among MC generators, very similar to those expected by \eposlhc\ and \qgsjetII;
whereas \sibyll~2.1 and \qgsjet~01\ generate about 50\% less. However, \eposlhc\ and \qgsjetII\ ground-level 
showers have about 35\% or 80\% more hadron energy fraction than predicted by \pythia~6, \sibyll~2.1, or \qgsjet~01 
(Fig.~\ref{fig:Ehad_vs_E}, right). 
That EAS simulated with \sibyll~2.1 feature less hadrons is not unexpected at first sight, given the lower particle 
multiplicity per pp collision predicted by this model, but this is in contradiction with the largest value 
predicted by \qgsjet~01 (Fig.~\ref{fig:sigma_pp_vs_sqrts}, top right). The underlying mechanism responsible 
of such differences requires further investigation.\\

As found for the \elm\ component, the fraction of shower energy 
carried by hadrons at ground is increasingly reduced, by absorption in the atmosphere, for decreasing angles of incidence.
The EAS simulated with \eposlhc\ and \qgsjetII\ feature higher hadron energy fractions at ground
for all zenith angles, and \pythia~6 (with and without heavy-quark production) have a very similar 
fraction and inclination-dependence as found with \qgsjet~01, whereas \sibyll~2.1 is clearly below the rest of the models.
\begin{figure}[htbp!]
\centering
\includegraphics[width=0.51\textwidth]{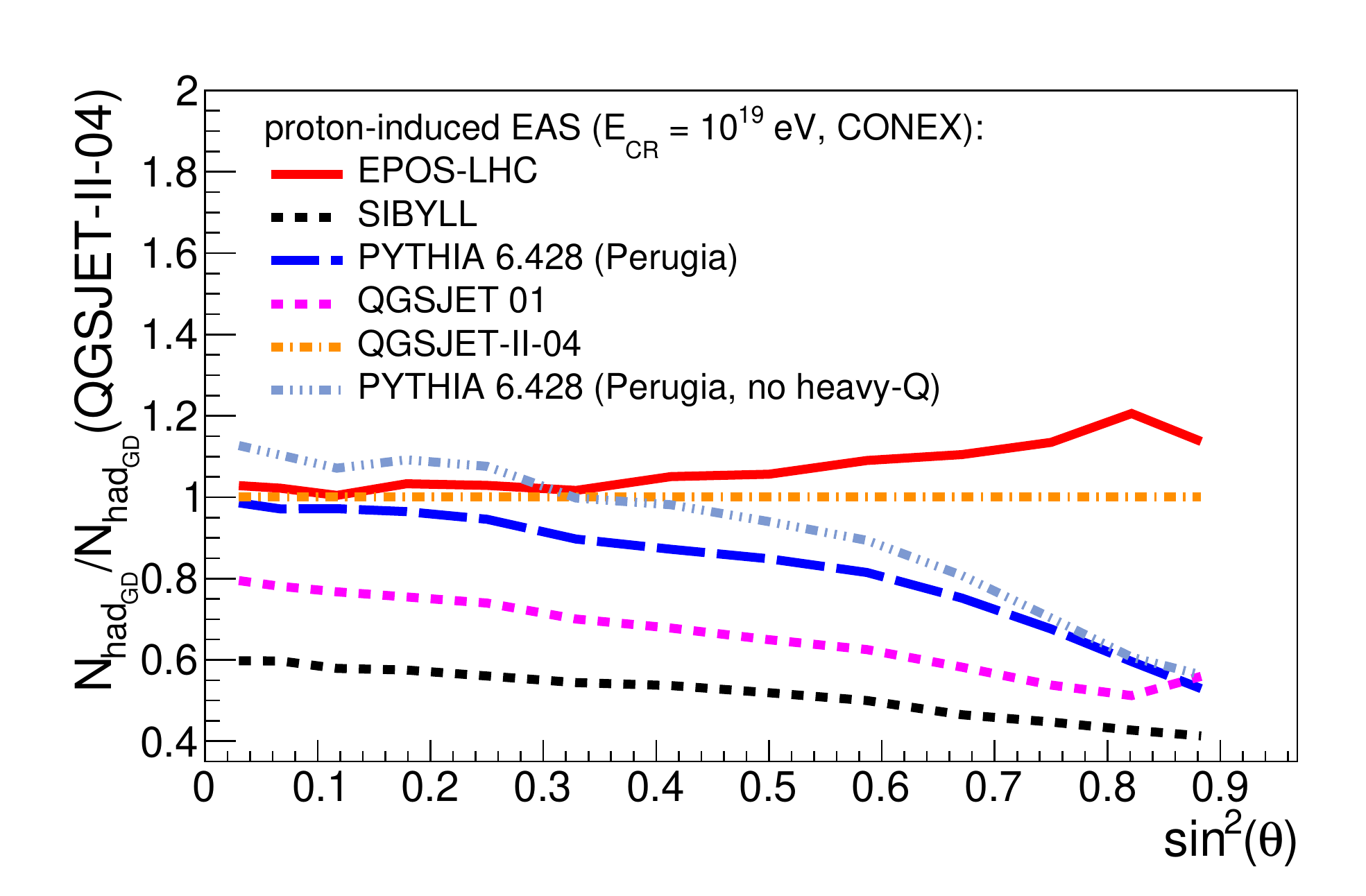}
\includegraphics[width=0.485\textwidth]{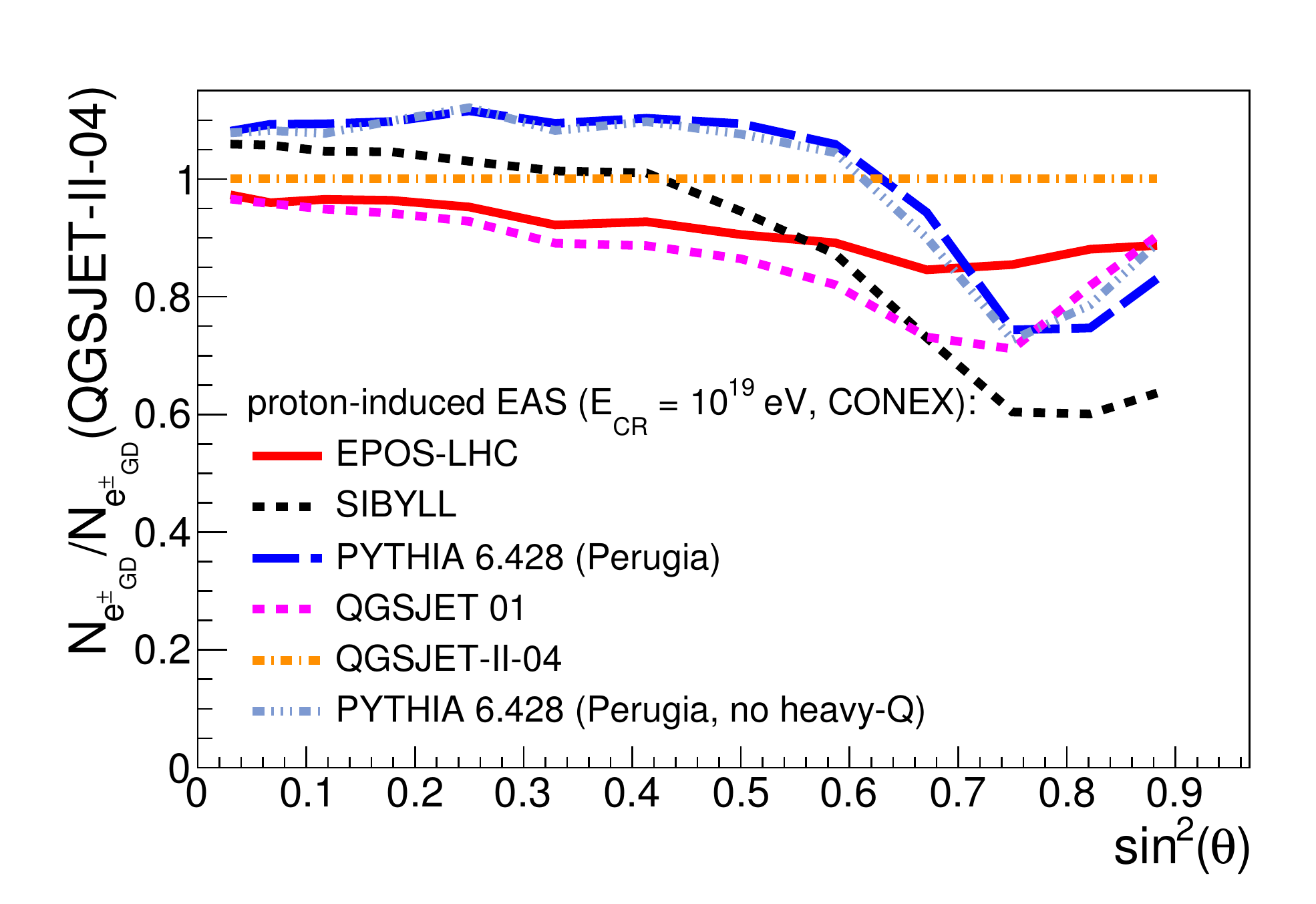}
\caption{Zenith-angle dependence of the number of hadrons (left) and electrons (right) at sea-level for proton-induced 
EAS of $\ECR = 10^{19}$~eV, predicted by the six MC event generators considered in this work {\it over} the \qgsjetII\ prediction.}  
\label{fig:emh_vs_sin2theta}
\end{figure}
This is better seen in Fig.~\ref{fig:emh_vs_sin2theta} (left), that compares the results of all the models with 
respect to the \qgsjetII\ prediction. The fast drop of the hadron production with zenith-angle seen for \pythia~6 showers, 
indicates that the hadronic component is absorbed relatively faster by the atmosphere for this MC generator than for the other models. 
The same phenomenon is observed for electrons in Fig.~\ref{fig:emh_vs_sin2theta} (right), except for very inclined showers
where an increase, related to the muonic component, is observed. Indeed, in that case, the electrons are coming from the 
decay of muons that are less attenuated by the atmosphere as shown in the next section.

\subsection{Muonic properties of proton-induced EAS}

\begin{figure}[htbp!]
\centering
\includegraphics[width=0.49\textwidth]{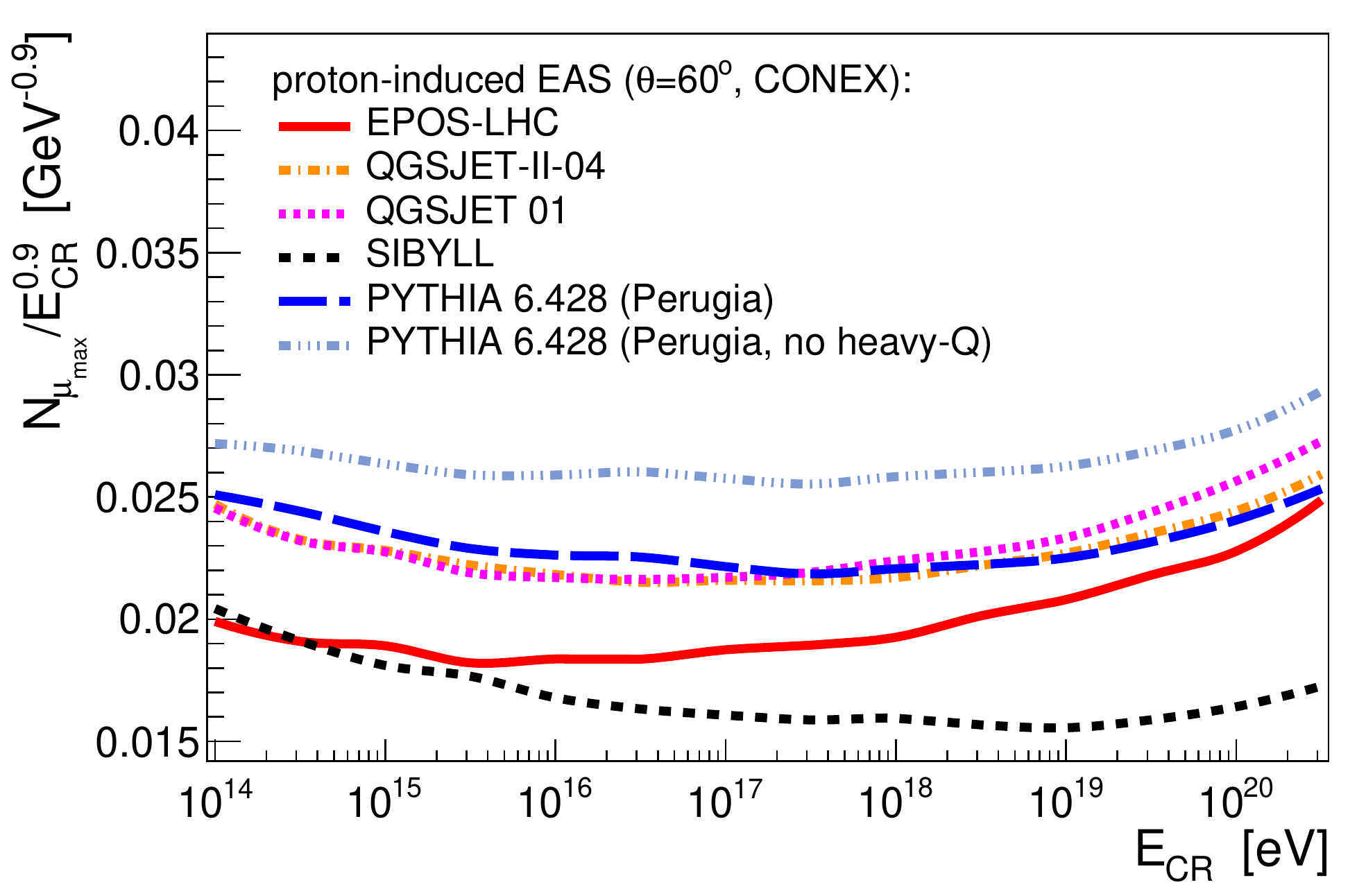}
\includegraphics[width=0.49\textwidth]{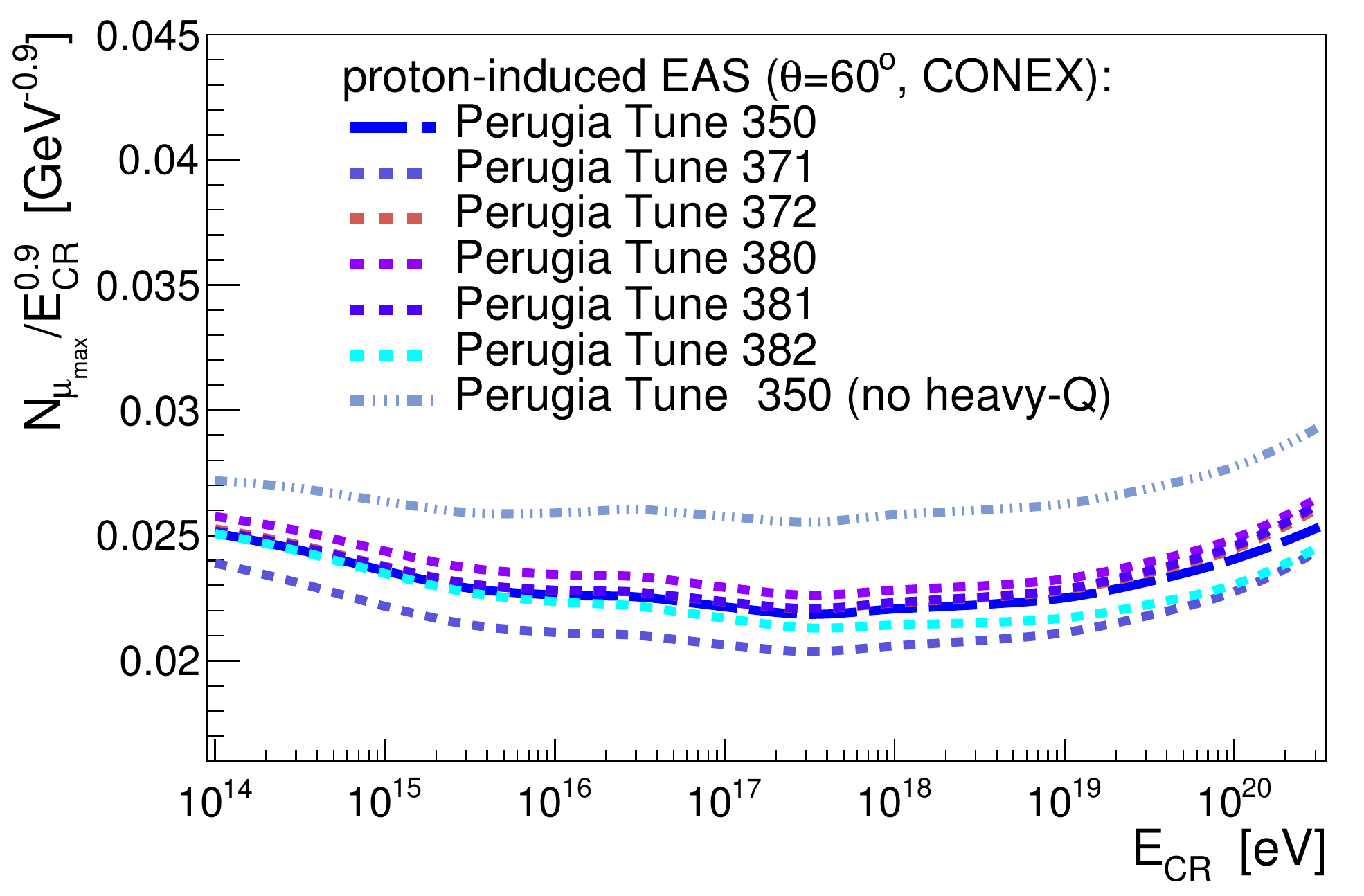}
\caption{Number of muons (normalized by $\ECR^{0.9}$) at shower maximum for inclined proton-induced showers 
($\theta = 60^\circ$) as a function of cosmic-ray energy, predicted by the six MC event generators (left), 
and for the seven \pythia~6 tunes (right) considered here.
\label{fig:N_mu_vs_E}}
\end{figure}

Whereas UHECRs detected straight from above the observatories have EAS dominated by their \elm\ component, 
photons and electrons are increasingly absorbed by the atmosphere in more inclined showers. At the same time, the larger
the column of air in non-vertical showers, the larger the number of charged hadrons 
that decay into muons. 
Thus, ultimately, the muon component dominates the shower composition for angles of arrival above $\theta\approx 45$--$50^\circ$. 
The study of the muon properties (number density, energy, distance from the shower axis) at ground-level in inclined EAS
provides thereby important insights on the hadronic development of the shower. 
Figure~\ref{fig:N_mu_vs_E} shows the number of muons at shower maximum (normalized by $\ECR^{0.9}$ to flatten out
the distribution according to the generalized Heitler model~\cite{Matthews:2005sd}) for inclined showers ($\theta = 60^\circ$) 
as a function of cosmic-ray energy for \pythia~6 and the RFT-based MC generators (left), and for the seven different
\pythia~6 tunes (right). The EAS generated with \pythia~6 produce about 
the same number of muons as those from \qgsjetII\ and \qgsjet~01, 
and 25--50\% more than \sibyll~2.1. Interestingly, whereas the different \pythia~6 tunes yield consistent number of muons within $\pm$5\%, running
\pythia~6 without heavy-quarks production generates $\sim$15\% more muons at shower maximum than all other MCs. This result 
indicates, first, that the main source of muons in \pythia~6 is clearly the decays of light-quark mesons (charged pions and kaons), 
and that heavy-quark production accounts for a negligible fraction of the total inclusive muons. Second, switching-off 
charm and bottom production seems to leave more room for $\pi^\pm$ and $K^\pm$ production, and thereby for an increased 
muon density in the showers. The main conclusion of this study is that, at least for a hydrogen atmosphere, there 
are ``non-exotic'' ways to increase by at least 15\% the muon density in the showers at ground. A similar conclusion has been reached 
with the latest version (2.3c) of \sibyll~\cite{Fedynitch:2018cbl}, that includes the production of heavy flavors and leads to a number of 
muons comparable to that of \epos\ and \qgsjetII~\cite{Pierog:2017nes}. Unfortunately, this latter MC generator was too recent 
to be tested with the modified atmosphere used in the current study.
In any case, one must be reminded that for real UHECR collisions on air, at variance with the result found for proton-hydrogen collisions,
\begin{figure}[htbp!]
\centering
\includegraphics[width=0.49\textwidth]{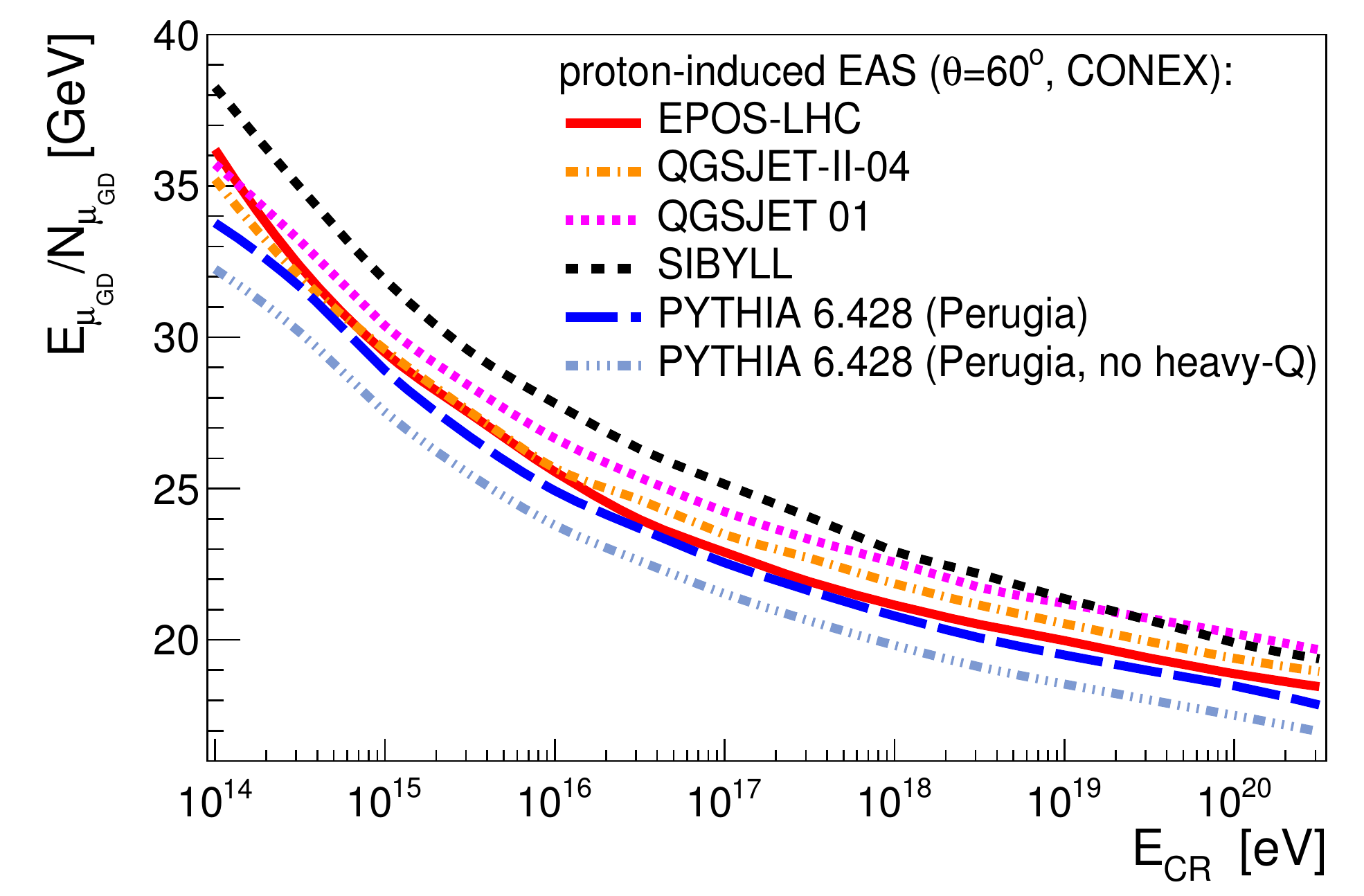}
\includegraphics[width=0.49\textwidth]{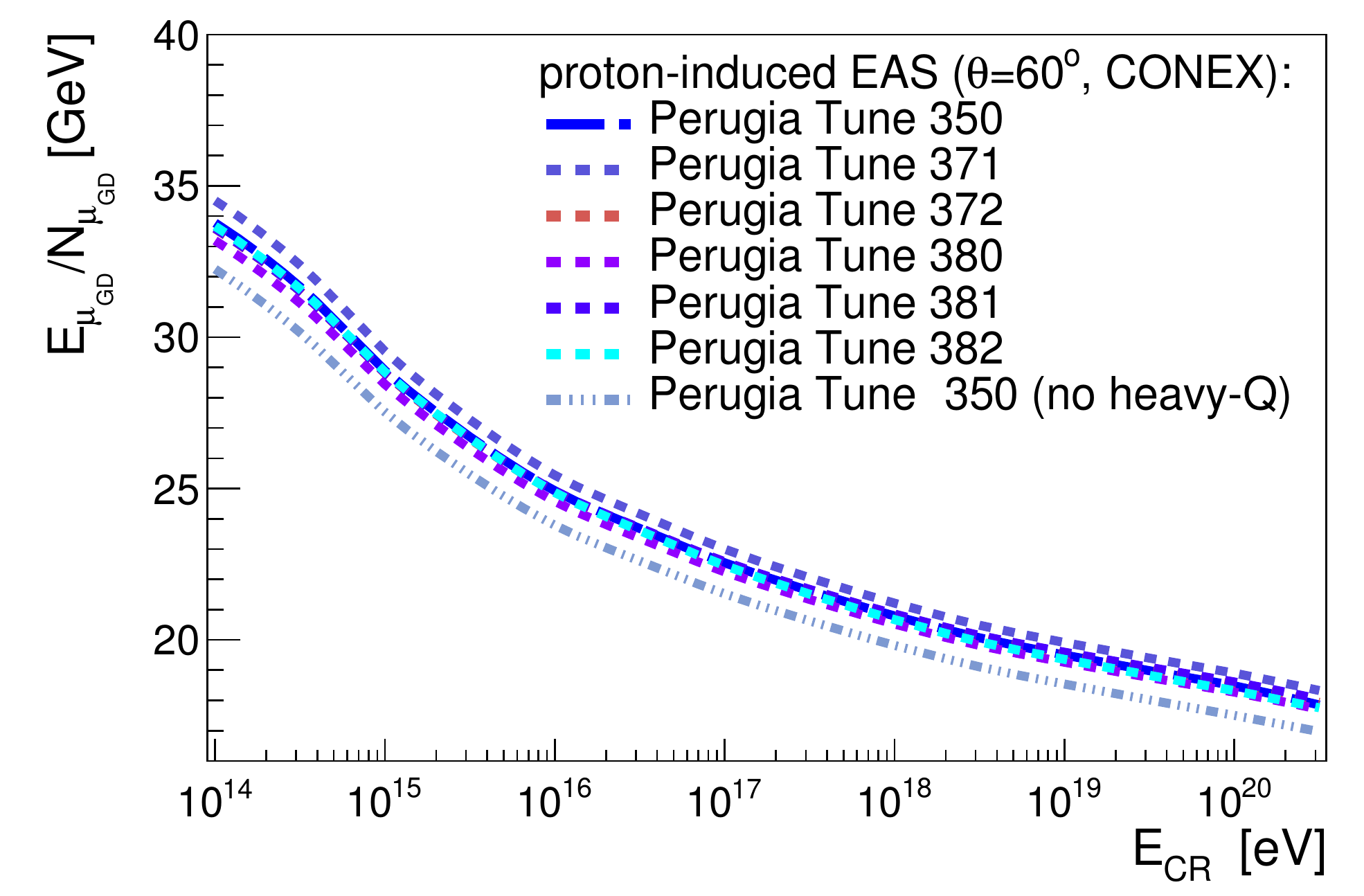}
\caption{Mean energy of muons at sea-level for inclined proton-induced showers ($\theta = 60^\circ$) 
as a function of cosmic-ray energy ($\ECR$) predicted by the six MC event generators (left), 
and for the seven \pythia~6 tunes (right) considered here.}  
\label{fig:Emu_vs_E}
\end{figure}
\eposlhc\ produces more $\mu^\pm$ than \qgsjetII\, with a similar $\ECR$-dependence~\cite{Pierog:2017nes}, while in Fig.~\ref{fig:N_mu_vs_E} 
(left) the slope of N$_\mu$ versus $\ECR$ predicted by \eposlhc\ is very different to that of all other models. Therefore, nuclear effects, 
absent in our current setup, definitely also play a role in the final inclusive production of muons measured in the data.\\

The average energy of the muons reaching sea-level for inclined showers ($\theta = 60^\circ$) 
is shown in Fig.~\ref{fig:Emu_vs_E} as a function of cosmic-ray energy for \pythia~6 and the RFT-based MC generators (left), 
and for the seven different \pythia~6 tunes (right). For such an observable, the features of the \pythia~6 proton-induced
showers are below the rest of the models: \pythia~6 predicts $\sim$5\% less energy per muon at ground than \qgsjetII\ and up to 
$\sim$10\% less than \qgsjet~01 or \sibyll~2.1, and only slightly below \eposlhc.
The right panel of Fig.~\ref{fig:Emu_vs_E} indicates small differences among \pythia~6 tunes,
although without heavy-quarks, \pythia\ produces more than 5\% less energetic muons. This is easily understood by the fact 
that high-energy $\mu^\pm$ coming from heavy-flavor meson decays are replaced by many more low energy muons produced after a long 
chain of hadronic interactions.\\

\begin{figure}[htbp!]
\centering
\includegraphics[width=0.49\textwidth]{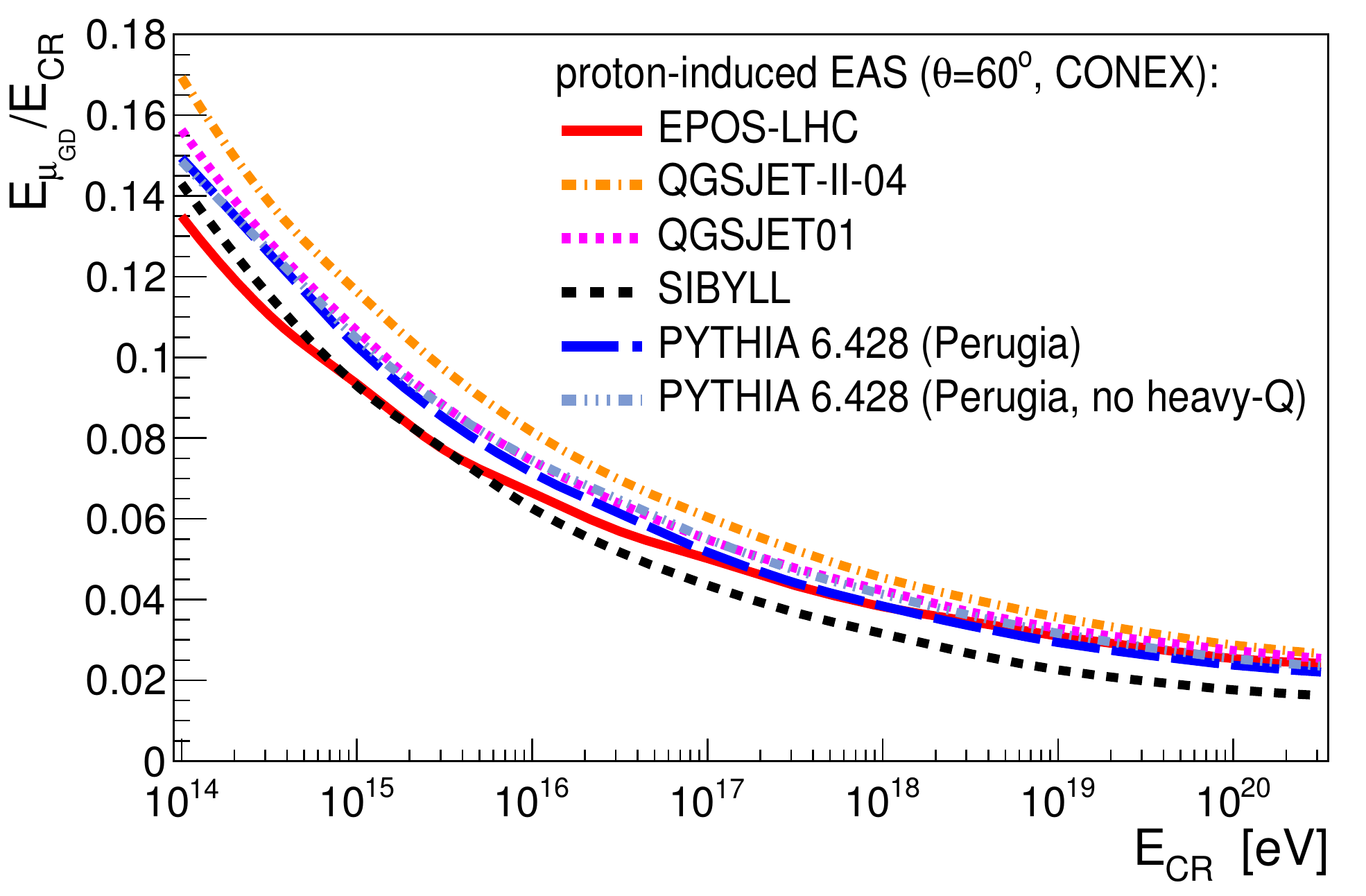}
\includegraphics[width=0.49\textwidth]{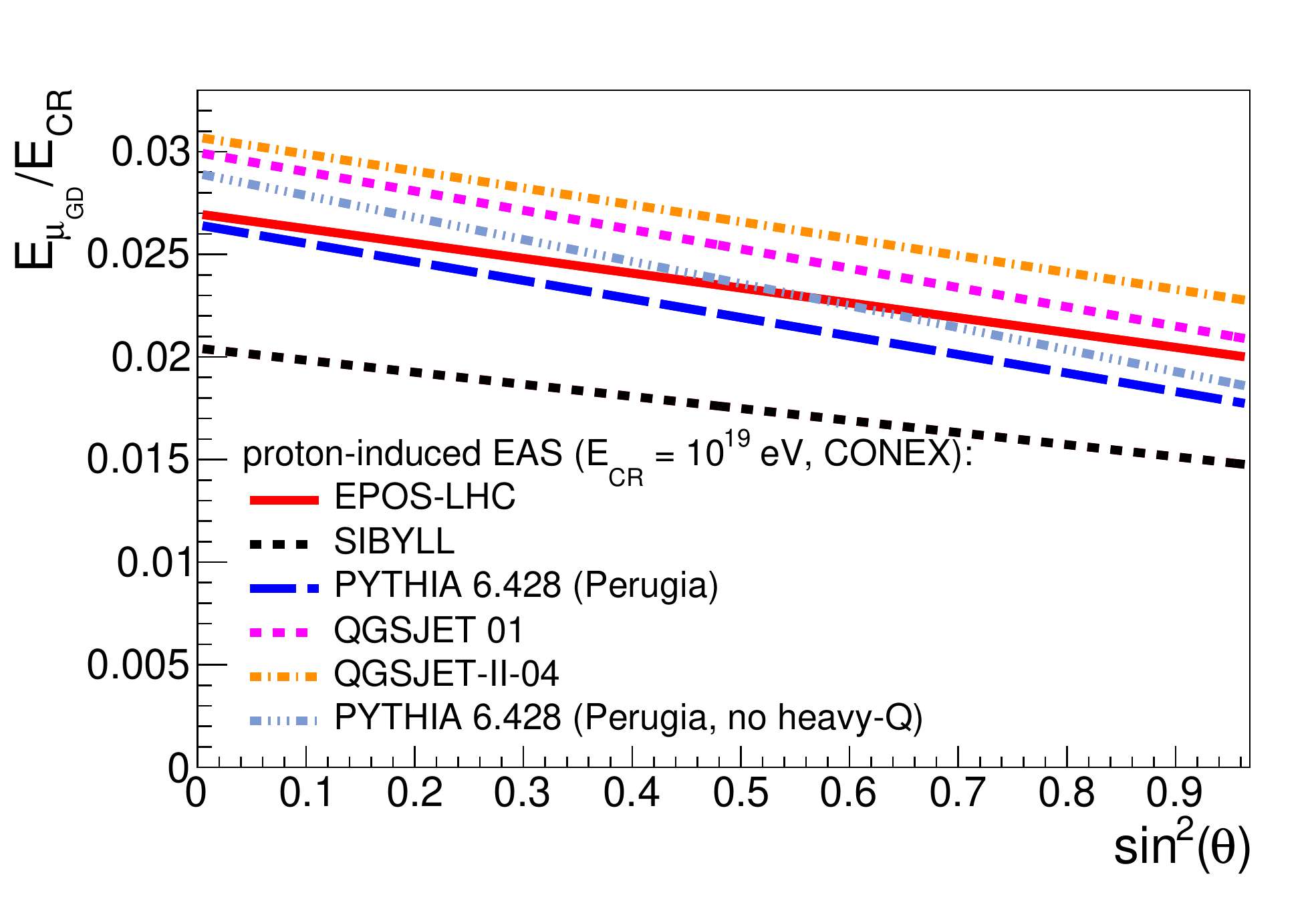}
\caption{Fraction of the CR energy carried by muons at sea-level for inclined proton-induced showers ($\theta = 60^\circ$) 
as a function of cosmic-ray energy ($\ECR$) (left), and  as a function of the (squared-sine) zenith angle of the incoming 
cosmic ray with $\ECR\approx$~10$^{19}$~eV (right), predicted by the six MC event generators considered in this work.}  
\label{fig:Emu_vs_sin2theta}
\end{figure}

The energy (left) and zenith-angle (right) dependence of the fraction of total energy carried by muons on ground for cosmic rays with $\ECR = 10^{19}$~eV 
are shown in Fig.~\ref{fig:Emu_vs_sin2theta}. The fraction of UHECR energy carried by muons is much less dependent on zenith angle than 
for $e^\pm$ and charged hadrons (right panels of Figs.~\ref{fig:Eem_vs_E} and~\ref{fig:Ehad_vs_E}). A similar hierarchy of MC
generators is found for the $\ECR$- (left) and for the $\theta$- (right) dependent results. The \qgsjetII\ and \qgsjet~01 simulations 
feature the largest fraction of energy carried by muons at sea-level, followed by \pythia~6 without heavy-quark production, \eposlhc, 
\pythia~6 (tune 350), and \sibyll~2.1. A slightly smaller attenuation length (steeper slope) is observed in the zenith-angle dependence of \pythia~6,
irrespectively of heavy flavor production, compared to all other models. This is consistent with the lower mean energy of the muons observed 
in Fig.~\ref{fig:Emu_vs_E}, and with the attenuation of the hadron component shown in Fig.~\ref{fig:emh_vs_sin2theta}.\\

\begin{figure}[htbp!]
\centering
\includegraphics[width=0.49\textwidth]{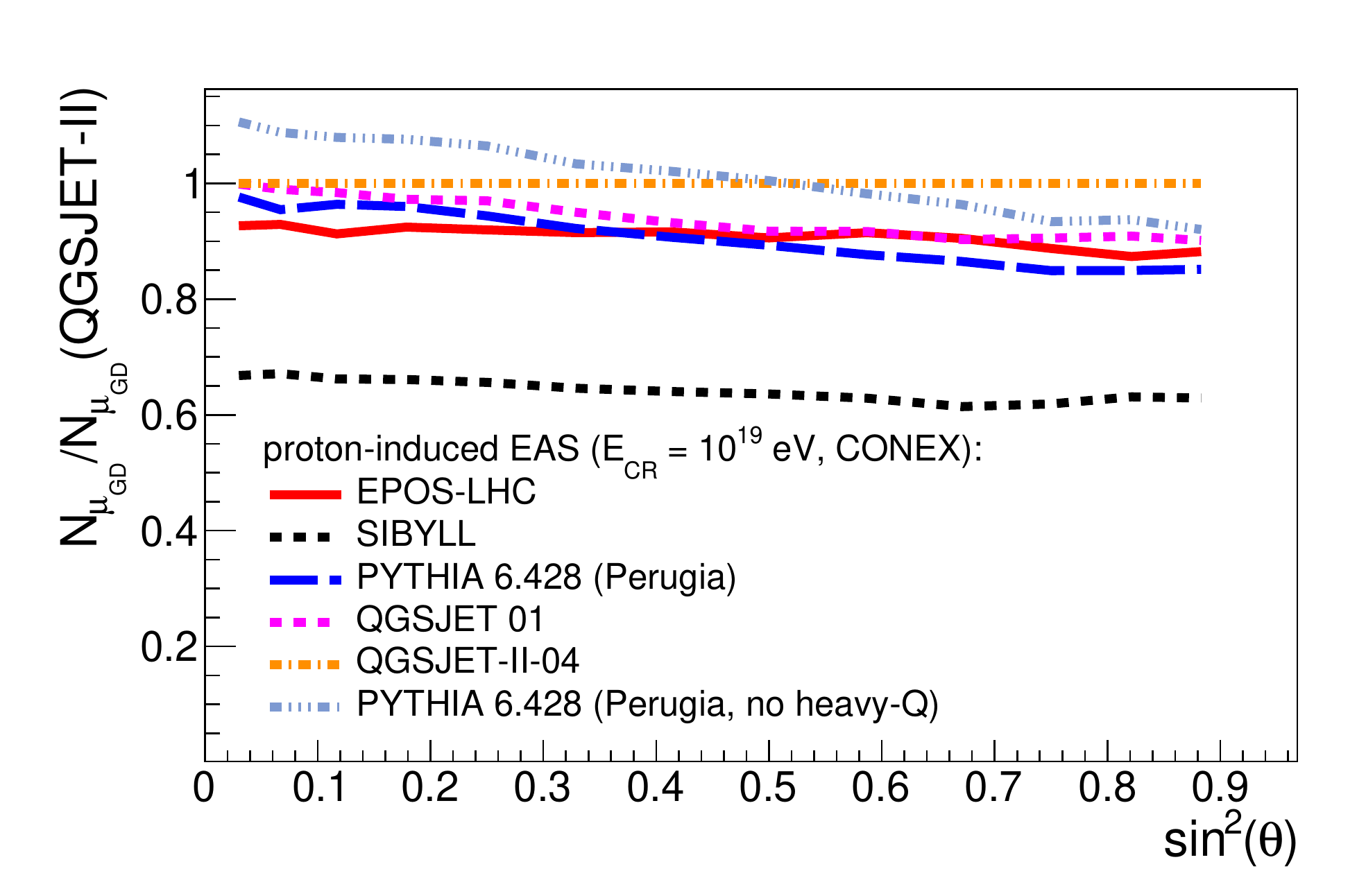}
\includegraphics[width=0.49\textwidth]{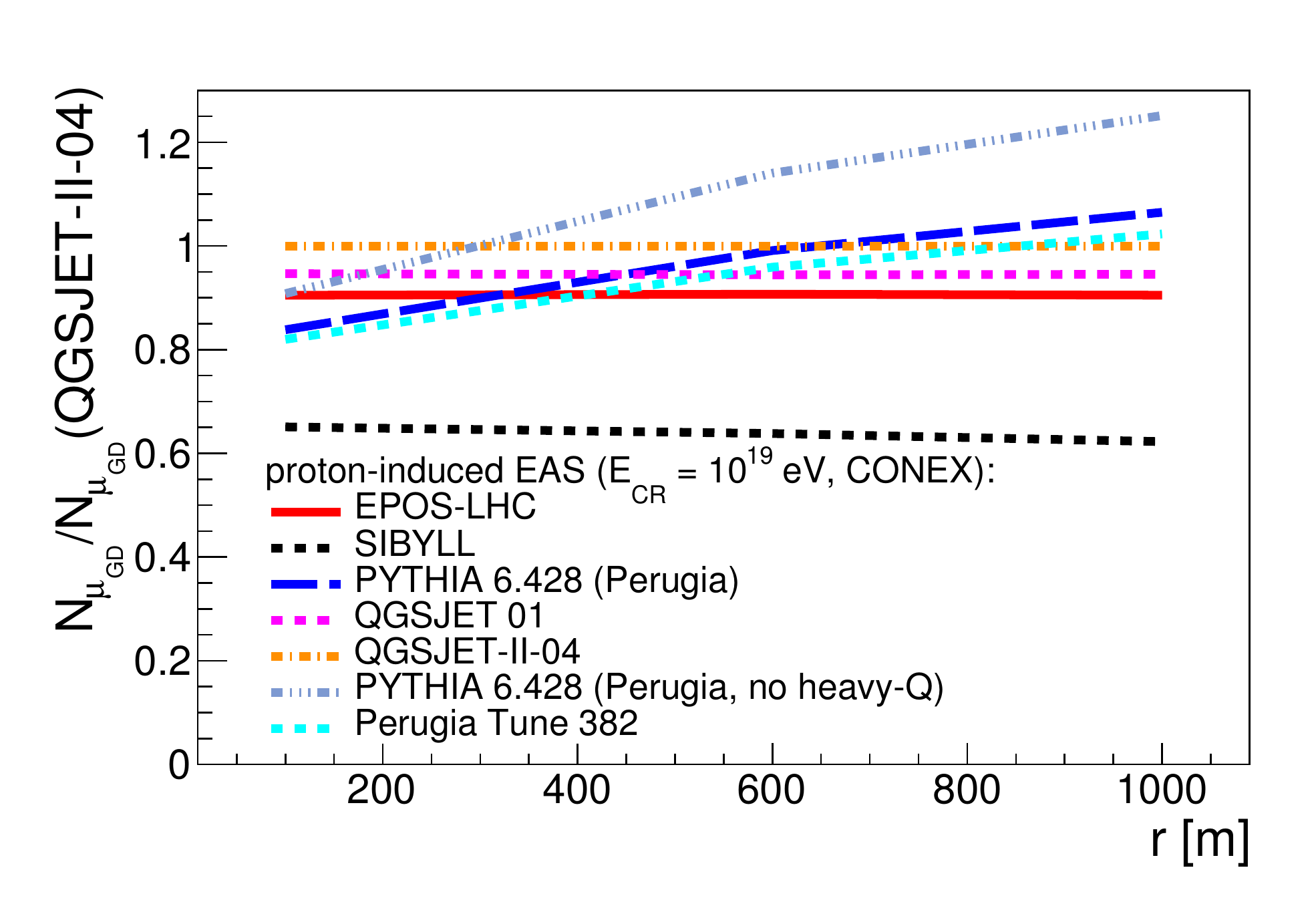}
\caption{Zenith-angle (left) and radial (distance to the shower core) (right) dependence of 
the number of muons at sea-level for proton-induced EAS of $\ECR = 10^{19}$~eV predicted by the six MC models 
considered here {\it over} the \qgsjetII\ prediction.}  
\label{fig:mu_vs_zenith}
\end{figure}

To further study the muon properties of the generated EAS, Fig.~\ref{fig:mu_vs_zenith} (left) shows the zenith-angle dependence 
(left) and the radial (distance to the shower core) dependence (right) of the number of muons at sea-level predicted by the 
different MC simulations {\it over} the \qgsjetII\ prediction (which is used as a reference here and features 
the largest muon density among all RFT-models), for proton-induced EAS of $\ECR = 10^{19}$~eV. 
Because of their lower average energy, the muons produced by \pythia~6 and \qgsjet~01 are absorbed by the atmosphere faster than for the other models.
Interestingly, Fig.~\ref{fig:mu_vs_zenith} (right) indicates that \pythia~6 features in general less muons than other MCs 
closer to the core shower (40--200~m) 
but predicts more muons 
for transverse distances larger than 600-m from the shower axis. The latter result is even more dramatic for \pythia~6 without heavy-quark production, which 
predicts up to 25\% more muons than \qgsjetII\ for very inclined showers. The higher density of muons at large radial
distances from the EAS core predicted by \pythia~6 is likely to be connected to the highest transverse momenta of the 
produced charged hadrons (that eventually decay into muons) in this pQCD-based Monte Carlo, as can be seen in the 
energy evolution of $\meanpt$ plotted in Fig.~\ref{fig:sigma_pp_vs_sqrts} (bottom left). Minijets, fragmenting into high-$\pT$ 
hadrons, are produced more copiously via multiparton interactions in \pythia~6 than in the rest of the MC event
generators. A possible way to test if such a mechanism is responsible for the larger number of muons at large transverse 
distances from the EAS axis, would be to modify the RFT-based models so that their fixed soft-hard $Q_0$ cutoff (which
controls the amount of pQCD minijets produced at a given $\sqrts$) is changed to a power-law running behavior 
as implemented in \pythia~6.

\section{Summary and conclusions}

A detailed study of the properties of extended air showers (EAS) generated by cosmic-ray protons
with energies $\ECR = 10^{14}$--$10^{20}$~eV has been carried out with a fast \conex\ simulation of
a hydrogen atmosphere with the same density as air. The use of an atmosphere with ``Jupiter-like'' 
composition allows one to interface the \pythia~6 Monte Carlo (MC) event generator, commonly used
in collider physics and tuned to reproduce the LHC proton-proton (pp) data, and compare its results to
those predicted by hadronic MC generators typically used in cosmic-ray studies (\eposlhc, \qgsjetII,
\qgsjet~01, and \sibyll). At variance with the latter hadronic models, based on Gribov's Reggeon Field Theory (RFT),
\pythia~6 contains factorized hard perturbative processes producing energetic QCD jets as well as charm and bottom quarks, that 
could potentially explain recent data--theory divergences, in particular, regarding the 
characteristics of the muons produced in EAS.\\

We have first studied the overall properties of pp collisions (inelastic cross section, charged-particle
multiplicity, mean transverse momentum, and inelasticity) as a function of cosmic-ray energy,
and found that all models reproduce the existing data up to equivalent energies of $\ECR\approx 10^{17}$~eV.
Beyond that energy, models start to deviate in their predictions up to $\ECR\approx 10^{20}$~eV, with \pythia~6
producing increasingly larger transverse momenta particles, and having overall lower pp inelasticity than
the rest of approaches.
We have then compared the properties of the generated proton-induced showers for key EAS variables 
such as the mean altitude of the shower maximum $\Xmax$ and the width of its associated 
fluctuations $\smax$. Detailed characteristics of the electromagnetic, charged-hadronic, and muonic
components of the EAS at the shower maximum and at sea-level have been studied as a function of 
the primary cosmic-ray energy, zenith-angle, and transverse distance from the shower axis.\\

The first generic conclusion reached is that, in general, the values and the $\ECR$-evolution of 
$\Xmax$ and $\smax$, as well as of the density and energy of $e^\pm$ and charged-hadrons at sea-level,
are quite similar in \pythia~6 and the standard cosmic-ray hadronic models. 
The second generic result is that changes in the \pythia~6 parameter settings (using different ``tunes'' of 
the semi-hard scattering and hadronization dynamics) result in very similar EAS properties, except when switching-off 
completely charm and bottom production, which leads to increased charged hadron and muon production, in particular at 
large transverse distances from the shower axis. The latter observation seems to indicate that heavy-quark production
(decaying into hard muons) is {\it not} the physical mechanism responsible of the overall increased $\mu^\pm$ production
observed in the data in comparison to the model predictions.\\

Looking into more detail, the \pythia~6 showers feature deeper penetration in the atmosphere (\ie\ larger 
shower maximum position) and smaller $\smax$  fluctuations, likely due to a reduced p-p inelasticity, 
compared to the other MC event generators (and similar to those predicted by \sibyll, although for very 
different underlying physical reasons). The characteristics of the electromagnetic 
component of proton-induced \pythia~6 EAS are found in-between those of other MC generators. The 
\pythia~6 hadron shower density at shower maximum is similar to that found with \eposlhc\ and \qgsjetII, 
whereas the corresponding fraction of shower energy carried by hadrons at sea-level is smaller than the one predicted
by the latter models and more similar to \sibyll\ or \qgsjet~01.\\


The properties of the muon component of the proton-induced EAS in \pythia~6 are significantly different 
than those predicted by the RFT-based hadronic models. First, in general \pythia~6 predicts 
a total muon density and energy at sea-level in between that of other MCs (more than \eposlhc\ but 
a bit less than \qgsjetII). However, switching-off charm and bottom production seems to leave more room 
for charged pion and kaon production, and thereby for a 15\% increased muon density in the showers. 
This result points out, first, that the main sources of muons in \pythia~6 are the decays of 
light-quark mesons (charged pions and kaons), and that heavy-quark production accounts for a negligible 
fraction of the inclusive muons. Secondly, this proves that, at least for a hydrogen atmosphere, there are 
``non-exotic'' ways to increase by at least 15\% the muon density in the shower at ground. However, one must be 
reminded that for real UHECR collisions on air, at variance with the results found in our proton-hydrogen collision setup, 
\eposlhc\ produces more $\mu^\pm$ than \qgsjetII. Therefore, nuclear effects, that are absent in our 
current setup, definitely also play a role in the final production of muons observed in the data~\cite{Pierog:2017nes}.\\

The most clear-cut result is that \pythia~6 (with or without heavy-quark production) features increasingly 
harder muon lateral distributions compared to the rest of models. \pythia~6 showers feature a bit less muons 
close to the core (40--200~m radial distance), but 10--30\% more at 600-m and at 1-km from the shower axis.
The higher density of muons at large radial distances predicted by \pythia~6 is likely to be connected to the 
highest transverse momenta of the produced charged hadrons in this model. The underlying physical reason
being that pQCD minijets, fragmenting into high-$\pT$ hadrons and eventually decaying 
into hard muons, are produced more copiously via multiparton interactions in \pythia~6 than in the rest 
of the MC event generators.\\

In summary, this is the first time, to our knowledge, that the \pythia\ event generator has been used to analyze ultrahigh-energy 
cosmic-rays showers. Our study indicates that retuning the production of multiple (hard) minijets 
in the standard cosmic-rays MC generators, combined with improved nuclear effects, will have a bigger 
impact on the muonic EAS properties, and on the potential resolution of several muon ``anomalies'' observed in UHECR showers, 
than incorporating the generation of charm and bottom quarks into the RFT-based models. The present work provides novel 
insights into the microscopic dynamics of hadronic collisions that are relevant for the understanding of the energy 
and identity of the highest-energy cosmic-rays in the universe.

\section*{Acknowledgments}

We are grateful to Awadallah~Mekki for help in preliminary studies of this topic.




\end{document}